\documentclass[aps,pre,twocolumn,superscriptaddress,nofootinbib]{revtex4-2}

\usepackage{amsmath,amssymb,bm,mathtools}
\usepackage{graphicx}
\usepackage{booktabs}
\usepackage{xcolor}
\usepackage[colorlinks=true,linkcolor=blue,citecolor=blue,urlcolor=blue]{hyperref}

\newcommand{\ii}{\mathrm{i}}

\newcommand{\avg}[1]{\left\langle #1\right\rangle}
\newcommand{\eps}{\varepsilon}

\begin{document}

\title{A solvable normal form for coupled swarmalators}

\author{Kevin P. O'Keeffe}
\affiliation{Starling Research Institute}

\date{\today}

  \begin{abstract}
  Swarmalators are mobile generalizations of phase oscillators. Introduced to model
  systems in which sync and
  self-assembly interact, they remain poorly understood theoretically.
  Unlike the Kuramoto model for coupled
  oscillators, existing swarmalator models lack a
  controlled microscopic derivation, and their ordered-state transitions remain only
  partly characterized. Here we address both. Building on Tanaka's reduction of chemotactic
  oscillators, we show that the one-dimensional
  swarmalator model -- previously introduced as an ad hoc toy model --
  is the leading-order phase--position approximation of intrinsically motile Tanaka-type
  chemotactic oscillators with contrast sensing and first-harmonic spatial transmission,
  in the weak-coupling, zero-lag limit. We then derive the exact phase-wave stationary
  neutral curve, characterize the sync and mixed transitions semi-analytically and numerically,
  show that the boundaries meet at a single cusp, show why a previously published Ott--Antonsen order-parameter formula misses the correlations of partially locked oscillators, and uncover a non-monotonic sync response absent in the Kuramoto model.
  \end{abstract}

\maketitle

\section{Introduction}
\begin{figure}[tbp]
\centering
\includegraphics[width=\columnwidth]{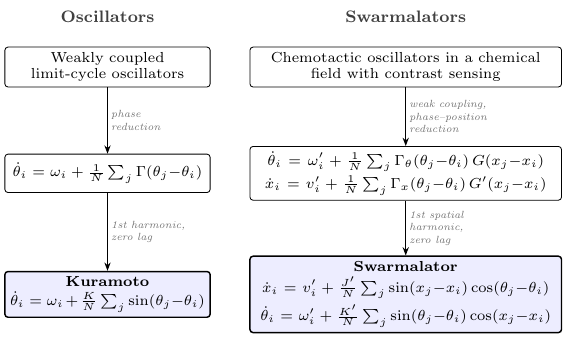}
\caption{Kuramoto and the swarmalator model have parallel origins. \emph{Left}:
weakly coupled limit-cycle oscillators reduce to a phase equation whose first harmonic,
with zero phase lag, is the Kuramoto model. \emph{Right}: the intrinsically motile
Tanaka-type realization of Sec.~III---with contrast sensing, first-harmonic spatial
transmission, and the temporal-response conditions stated there---reduces to a
phase--position equation whose weak-coupling, zero-lag limit is the 1D swarmalator model.}
\label{fig:reduction}
\end{figure}

  Synchronization theory has a canonical normal form: the
  Kuramoto model. It is the first-harmonic, zero-lag phase reduction
  of weakly coupled limit-cycle oscillators (Figure~\ref{fig:reduction}), and it organizes
  experiments from Josephson arrays to electrochemical
  oscillators \cite{Kuramoto1975,Strogatz2000,Acebron2005,Wiesenfeld1996}.
  Swarmalators are generalizations of oscillators that move around in
   space as well as synchronize in time \cite{OHS2017,OCP2022,Yoon2022}.
  Such phase--position coupling arises across many natural and engineered systems \cite{Suematsu2016,OHS2017,Heuthe2025},
  but a theoretical framework as systematic as Kuramoto's is lacking.

  The simplest tractable setting is nonidentical swarmalators
  on a 1D periodic domain \cite{Yoon2022}. The 1D ring model
  \eqref{eq:physical-x}--\eqref{eq:physical-th} has become the main theoretical
  testbed for swarmalator dynamics. It has been studied under thermal noise
  \cite{Hong2023ThermalNoise}, external forcing \cite{Anwar2024Forced1D}, time
  delay \cite{Blum2024Delay,OKeeffeHindes2026Delay}, pinning and driving
  \cite{Sar2023Pinning,Sar2024Driven}, attractive and repulsive interactions
  \cite{Hao2023Attractive}, higher-order couplings \cite{Anwar2024HigherOrder},
  pulsating dynamics \cite{Ghosh2025Pulsating}, and contrarian elements
  \cite{Sar2025Contrarian}. Alternative coupling architectures are also being explored,
  for example two-dimensional swarmalators with Winfree-type (pulse-like, state-dependent)
  phase coupling \cite{Acharya2025Winfree}, in place of the Kuramoto-type sine coupling
  used here; the reduction below realizes the Kuramoto-type architecture for a specified
  class of chemotactic oscillators. Yet in all these studies the model was treated as an ad
  hoc toy model, with no microscopic derivation; and although the incoherence threshold
  and several order parameters were known, its ordered-state neutral conditions and
  bifurcations remained incompletely characterized.

  Earlier work mapped which collective states
  exist and gave expressions for their order parameters, but left open
  the two questions that made Kuramoto a paradigm: does the model descend from a
  microscopic mechanism, and can its collective dynamics -- stabilities, bifurcations
  and so on -- be solved? This paper addresses both, within the scope stated below. Adapting Tanaka's reduction
  \cite{Tanaka2007}, we show that the model is the leading-order phase--position
  approximation of intrinsically motile Tanaka-type chemotactic oscillators with contrast
  sensing and first-harmonic spatial transmission, in the weak-coupling, zero-lag limit
  (Figure~\ref{fig:reduction}). We then derive the
  phase-wave neutral curve exactly, characterize the sync and mixed transitions using
  semi-analytical calculations and simulations, and show that the boundaries meet at a
  single cusp in parameter space.

\section{Model}
The 1D swarmalator model is 
\begin{align}
\dot x_i&=v_i'+\frac{J'}{N}\sum_{j=1}^N\sin(x_j-x_i)\cos(\theta_j-\theta_i),
\label{eq:physical-x}\\
\dot\theta_i&=\omega_i'+\frac{K'}{N}\sum_{j=1}^N\sin(\theta_j-\theta_i)\cos(x_j-x_i),
\label{eq:physical-th}
\end{align}
where $x_i,\theta_i\in S^1$ are the position and phase of the $i$-th swarmalator,
$(v_i', \omega_i')$ are natural frequencies, and $(J',K')$ the coupling strengths.
Eq.~\eqref{eq:physical-th} models synchronization (the $\sin$ term) that depends on
distance (the $\cos$ term). Eq.~\eqref{eq:physical-x} models the reverse: aggregation that
depends on phase similarity. We take $\omega_i'$ and $v_i'$ to be independent and Cauchy
distributed, with zero location parameter and common width $\Delta$, which we set to $1$ by
rescaling time. Sweeps at $J=1$ and $J=3$ show the same qualitative state sequence for four
other full-support distributions (Gaussian, Laplace, logistic, hyperbolic secant;
Supplemental Material).

\begin{figure}[tbp]
\centering
\includegraphics[width=0.99\columnwidth]{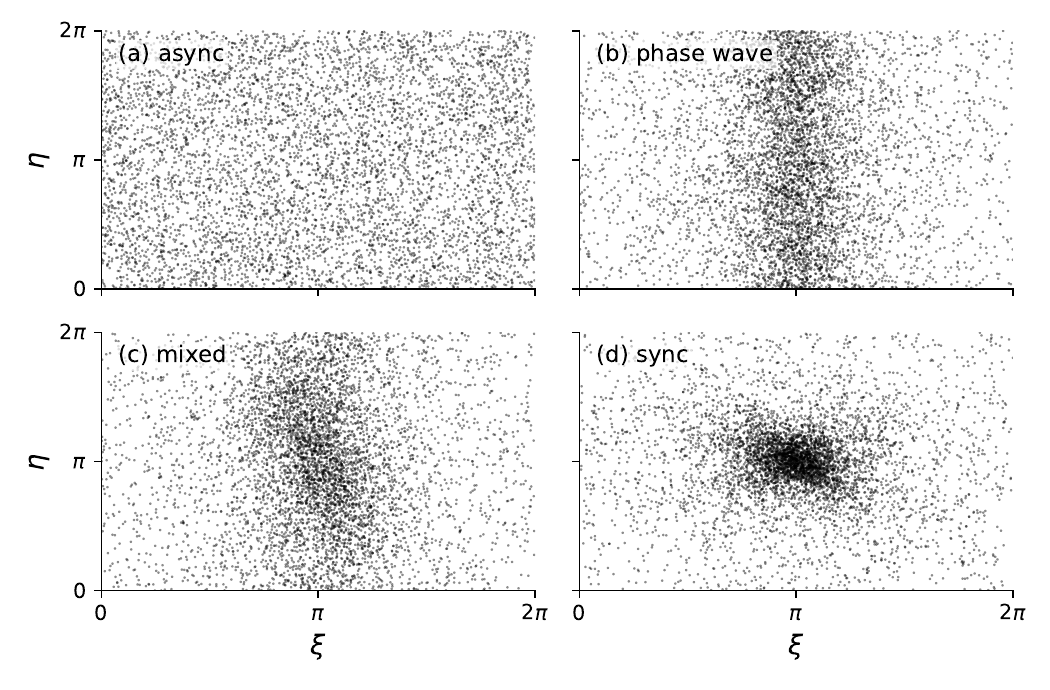}
\caption{Scatter plots of the four states in the
$(\xi,\eta)$ plane (black points), with the order-parameter conditions of each state listed
below. Direct simulation of
Eqs.~\eqref{eq:xi}--\eqref{eq:eta} with $N\approx10^7$ swarmalators ($3163^2$
deterministic Cauchy quantile nodes, $\Delta=1$), RK4 with $\Delta t=0.05$ to $t=140$ from
uniform random initial conditions; $6000$ randomly chosen swarmalators are shown.
$(K,J)=(3,2),(6,5),(6,4),(6,1)$ for (a)--(d).
(a) async $(0,0)$: uniform on the torus.
(b) phase wave $(r,0)$ [equivalently $(0,s)$ under $\xi\leftrightarrow\eta$]: one rainbow
coordinate locks, a single band.
(c) mixed $(r\!\ne\!s)$: both partially lock.
(d) sync $(r,r)$: both coordinates partially order equally, a compact cluster.}
\label{fig:gallery}
\end{figure}

\section{Chemotactic phase--position reduction}
Tanaka \cite{Tanaka2007} considered limit-cycle elements with internal state $X_i$ and
position $r_i$, coupled through a secreted chemical $S$:
$\dot X_i=f(X_i)+k\,g(S(r_i,t))$, $m\ddot r_i=-\gamma\dot r_i+\beta(X_i)\nabla S(r_i,t)$ and
$\tau\partial_tS=-S+d\nabla^2S+\sum_jh(X_j)\,\delta(r-r_j)$. Two features matter below. The
chemotactic sensitivity $\beta(X_i)$ depends on the internal state; since the field
oscillates at the oscillator frequency, a constant sensitivity would give no net drift over
a cycle. And the elements have no intrinsic motility.

On a ring this model does not by itself produce the swarmalator coupling, for two reasons.
First, the uniform chemical mode gives a position-independent phase coupling. Second,
point sources do not select the first spatial harmonic: the ring kernel has Fourier
coefficients $g_n\propto(1+dn^2+\ii\Omega\tau)^{-1}$, so $|g_2/g_1|\ge\tfrac14$ for all
parameters (Supplemental Material). We therefore study a \emph{Tanaka-type} extension with
three added ingredients: contrast sensing (the oscillators respond to $S-\bar S$, with
$\bar S$ the spatial mean), first-harmonic spatial transmission, and an intrinsic motility
$v_i'$. A minimal explicit realization uses isochronous Stuart--Landau oscillators
$X_i=(u_i,w_i)$ with $\Omega_i=\Omega+\omega_i'$,
\begin{align}
\dot u_i&=\lambda(1-u_i^2-w_i^2)u_i-\Omega_iw_i+a\,C(x_i,t),\label{eq:tanaka-X}\\
\dot w_i&=\Omega_iu_i+\lambda(1-u_i^2-w_i^2)w_i,\nonumber\\
\dot x_i&=v_i'+b\,u_i\,\partial_xS(x_i,t),\label{eq:tanaka-R}\\
\tau\partial_tS&=-S+d\,\partial_x^2S+\frac{q}{N}\sum_ju_j\big[1+\cos(x-x_j)\big],
\label{eq:tanaka-S}
\end{align}
with contrast $C=S-\bar S$. Here $a$ is the chemical response of the oscillator, $bu_i$ the
state-dependent sensitivity (divided by friction), and $q\,u_j$ the secretion, spread over
the unit-radius ring with the profile $1+\cos(x-x_j)$. For initial fields in the span of
$1$, $\cos x$ and $\sin x$, the field stays exactly in three modes, $S=S_0+S_c\cos x+S_s\sin x$;
other initial modes $n\ge2$ decay as $e^{-(1+dn^2)t/\tau}$. Contrast sensing removes $S_0$
from the oscillators, and it never enters the motion. (The symbols $S$, $H$, $a$, $b$, $q$
and $\lambda$ are local to this section.)

Take $q>0$ without loss of generality (its sign can be absorbed into $a$ and $b$). For
weak coupling---$a=\eps_{\rm c}a_0$, $b=\eps_{\rm c}b_0$, $v_i'=O(\eps_{\rm c})$ and
$\omega_i'=O(\eps_{\rm c})$, with $\eps_{\rm c}\ll1$ and $q$, $a_0$, $b_0$ of order one relative to
$\Omega$ and $\lambda$---the amplitudes $\rho_i$ in $(u_i,w_i)=\rho_i(\cos\phi_i,\sin\phi_i)$ stay near $1$,
positions and phase offsets $\theta_i=\phi_i-\Omega t$ are slow, and the first spatial mode responds to secretion at
frequency $\Omega$ with the complex gain $H=q/(1+d+\ii\Omega\tau)$. Averaging over one
cycle, with $\langle\cos\phi_i\cos\phi_j\rangle=\tfrac12\cos(\theta_j-\theta_i)$ and
$\langle-\sin\phi_i\cos\phi_j\rangle=\tfrac12\sin(\theta_j-\theta_i)$, gives
\begin{align}
\dot x_i&=v_i'+\frac{J'}{N}\sum_j\sin(x_j-x_i)\cos(\theta_j-\theta_i-\beta_L),
\label{eq:ring-x}\\
\dot\theta_i&=\omega_i'+\frac{K'}{N}\sum_j\sin(\theta_j-\theta_i+\alpha_L)\cos(x_j-x_i),
\label{eq:ring-th}
\end{align}
with $J'=bq/(2L)$, $K'=aq/(2L)$, $L=\sqrt{(1+d)^2+(\Omega\tau)^2}$, and chemical lags
$\beta_L=-\alpha_L=\delta$, where
$\delta=\arctan[\Omega\tau/(1+d)]$. For fast chemistry, $\Omega\tau/(1+d)\ll1$, the lags
vanish and Eqs.~\eqref{eq:ring-x}--\eqref{eq:ring-th} become
Eqs.~\eqref{eq:physical-x}--\eqref{eq:physical-th} with $J'=bq/[2(1+d)]$ and
$K'=aq/[2(1+d)]$. The signs of $a$ and $b$ are free, so $J'$ and $K'$ can have either
sign. Alternatively, rotating the oscillator's chemical input and sensitivity by $\delta$
cancels the lag at finite $\tau$ (Supplemental Material).

The claim is therefore precise: \emph{the 1D swarmalator model is the leading-order
phase--position approximation of intrinsically motile Tanaka-type chemotactic oscillators
with contrast sensing and first-harmonic spatial transmission, in the weak-coupling,
zero-lag limit.} We call the model a normal form only in this restricted sense: the leading-order reduction
within the stated class, not a generic limit of all chemotactic oscillators. ``Leading
order'' has a definite meaning. For a fixed finite population,
the microscopic positions and phases track the averaged model with error $O(\eps_{\rm c})$ over
times $t\le T/\eps_{\rm c}$ for any fixed $T$, where $\eps_{\rm c}$ is the coupling scale. Direct
integration of Eqs.~\eqref{eq:tanaka-X}--\eqref{eq:tanaka-S} confirms this first-order
convergence; omitting the lag, the uniform mode, or a second spatial harmonic instead
leaves an error that does not shrink (Supplemental Material, Sec.~S1). Three limitations
should be kept in view. Contrast sensing and first-harmonic transmission are modeling
choices, not consequences of Tanaka's point-source model. The natural velocities $v_i'$
are added, not derived. And the Cauchy disorder used below, with unbounded tails, is a
statistical idealization lying outside the finite-population, bounded-detuning estimate.
For a general limit cycle, first-harmonic transmission and zero chemical lag are not enough:
the secretion, sensitivity and phase-response waveforms must also satisfy temporal
correlation conditions, which the Supplemental Material gives (including the effect of
shear). The explicit realization above satisfies them.

\begin{figure}[tbp]
\centering
\includegraphics[width=0.9\columnwidth]{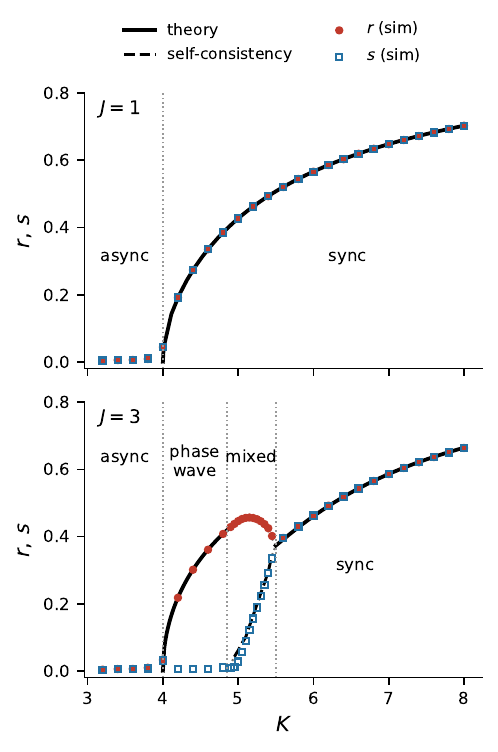}
\caption{Order parameters $r$ (red circles) and $s$ (blue squares) versus coupling $K$. Curves
are theory only; markers are direct simulation only. Solid black: exact phase-wave amplitude
$r_{\rm pw}=\sqrt{1-4/K}$ and the semi-analytical sync amplitude \eqref{eq:rsync}. Dashed
black: numerical mean-field self-consistency \eqref{eq:mixed} for the mixed branch (Sec.~IV.C),
computed at $4.95\le K\le5.45$. Its connections to the exact phase-wave endpoint
$K_2(3)=4.855$ and to an estimated merge with sync at $K\approx5.5$ are guides to the eye,
not computed roots.
Markers: direct simulation of Eqs.~\eqref{eq:xi}--\eqref{eq:eta}, $N=2.5\times10^5$ ($500^2$ deterministic Cauchy quantile nodes, $\Delta=1$), RK4 with $\Delta t=0.05$, adiabatic up-sweep in $K$ from $3.2$ to $8.0$ in steps of $0.2$ from uniform random phases, relaxation $t=200$ and averaging $t=120$ at each $K$. For
$4.9\le K\le5.45$ at $J=3$, independent long runs at the same $N$ (relaxation $t=1500$,
averaging $t=500$) are used instead. Dotted lines mark the
transitions. \emph{Top}, $J=1<J_c=2$: a single async$\to$sync
transition at $K=4$ with $r=s$. \emph{Bottom}, $J=3>J_c=2$: the full
async$\to$phase wave$\to$mixed$\to$sync cascade.}
\label{fig:rs}
\end{figure}
\section{Analysis} In sum/difference coordinates 
$\xi_i=x_i+\theta_i$, $\eta_i=x_i-\theta_i$, the model becomes especially clean
\begin{align}
\dot\xi_i&=\nu_i-Kr\sin\xi_i-Js\sin\eta_i,\label{eq:xi}\\
\dot\eta_i&=\mu_i-Jr\sin\xi_i-Ks\sin\eta_i.\label{eq:eta}
\end{align}
Here $(\nu_i,\mu_i)=(v_i'+\omega_i', v_i'-\omega_i')$\footnote{The transformed marginals are Lorentzian of half-width $2$, correlated through the same underlying $v',\omega'$.},
$(K,J)=((J'+K')/2, (J'-K')/2)$, and the order parameters are
\begin{equation}
re^{\ii\phi}=\frac1N\sum_j e^{\ii\xi_j},\qquad
se^{\ii\psi}=\frac1N\sum_j e^{\ii\eta_j}.
\label{eq:Z}
\end{equation}
For the stationary states considered here, constant phase shifts set $\phi=\psi=0$.

\emph{Physical meaning of $r$ and $s$.} Since $e^{\ii\xi}=e^{\ii x}e^{\ii\theta}$ and
$e^{\ii\eta}=e^{\ii x}e^{-\ii\theta}$, $r$ and $s$ measure \emph{position--phase
correlations} rather than aggregation or synchrony separately. $r=1$ means
$x_j+\theta_j$ is the same for every swarmalator, so phase winds backwards around the ring
(a ``rainbow'' of one handedness); $s=1$ is the opposite handedness. Two limits fix
intuition. If positions and phases were statistically independent, with
$Z_x=\avg{e^{\ii x}}$ and $Z_\theta=\avg{e^{\ii\theta}}$, then $r=s=|Z_x||Z_\theta|$:
$r=s>0$ requires spatial clustering \emph{and} phase coherence. Conversely, if $x$ and
$\theta$ are each uniform but perfectly correlated ($\theta_j=-x_j+c$), then $r=1$, $s=0$:
no aggregation and no global synchrony, yet complete order. More generally, when the disorder
law obeys $g(v',\omega')=g(v',-\omega')$, as for independent centered Cauchy disorder, the
model is invariant under the reflection $(\theta,\omega')\to(-\theta,-\omega')$, which
exchanges $\xi\leftrightarrow\eta$, $\nu\leftrightarrow\mu$ and $r\leftrightarrow s$.
Reflection-symmetric states have $r=s$; the converse need not hold, since equality of two
moments does not fix the whole density. States with $r\neq s$ break the symmetry and carry
a net handedness. Sync is the symmetric ordered branch, $r=s>0$, coherent in both rainbow
coordinates. Because $(x,\theta)\mapsto(\xi,\eta)$ is two-to-one, $(x,\theta)$ and
$(x+\pi,\theta+\pi)$ have the same $(\xi,\eta)$, so a locked cluster can lift to two antipodal
clusters in the original variables; $r$ and $s$ do not fix their relative populations or the
ordinary moments $Z_x,Z_\theta$. Both the phase wave and the mixed state break the
reflection: in the phase wave one rainbow amplitude vanishes, and in the mixed state both
are nonzero and unequal. Table~\ref{tab:states} summarizes.

\begin{table}[tbp]
\caption{The four collective states. $K_2(J)$ is the phase-wave neutral curve and $K_3(J)$ the
sync$\to$mixed transition (Fig.~\ref{fig:phase-diagram}). Ranges describe the sector $K,J\ge0$
with $J>J_c=2$; for $0\le J<2$ only async and sync occur, with the transition at $K=4$.}
\label{tab:states}
\begin{ruledtabular}
\begin{tabular}{llll}
State & $(r,s)$ & Picture & Range\\
\hline
Async & $r=s=0$ & $x,\theta$ uniform & $K<4$\\
Phase wave & $r>0=s$ & $x,\theta$ uniform, & $4<K<K_2$\\
 & \ (or $r\!\leftrightarrow\! s$) & \ $\theta\approx\mp x+c$ & \\
Mixed & $r,s>0$, $r\neq s$ & partial clusters, & $K_2<K<K_3$\\
 & & \ net handedness & \\
Sync & $r=s>0$ & coherent in $\xi,\eta$$^{a}$ & $K>K_3$\\
\end{tabular}
\end{ruledtabular}
\raggedright\footnotesize $^{a}$In $(x,\theta)$, one cluster or two antipodal clusters related
by $(x,\theta)\to(x+\pi,\theta+\pi)$.
\end{table}

Eqs.~\eqref{eq:xi}--\eqref{eq:eta} have nice structure. They are a pair of coupled Kuramoto models that fully
decouple at the limit point $J=0$. Increasing $J$ from zero thus allows us to tune the model
from the ``Kuramoto regime'' into the swarmalator regime about which much less is known.

\begin{figure}[tbp]
\centering
\includegraphics[width=0.92\columnwidth]{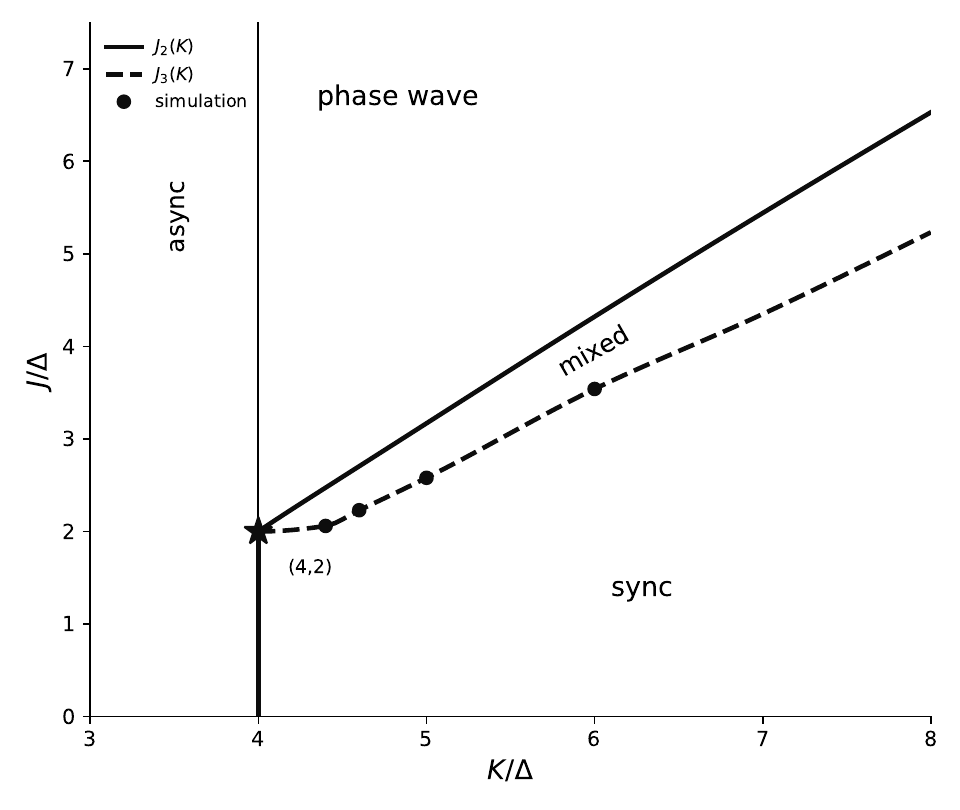}
\caption{Phase diagram in the $(K/\Delta,J/\Delta)$ plane. Solid curve: exact
phase-wave stationary neutral curve $J_2(K)$ from $F(K,J)=0$, consistent with the simulated
transitions. Circles: sync$\to$mixed onsets from
adiabatic up-sweeps in $J$ at fixed $K$, started from sync ($64^2$--$72^2$ deterministic
Cauchy nodes, $\Delta=1$, RK4 with $\Delta t=0.05$, relaxation $150$--$180$ and averaging
$80$--$100$ per step, $\Delta J=0.025$--$0.08$), located by a square-root fit of $(r-s)^2$. Dashed curve: sync boundary $J_3(K)$ drawn as a monotone (PCHIP)
interpolation through the cusp, the circles, and the large-$J$ asymptotic
$K_3=J+\tfrac{2}{\pi}\ln(2J)+\tfrac{4}{\pi}$ [Eq.~\eqref{eq:Kc3}] at $K\approx7$--$9$; it is a
guide to the eye, not an independent prediction, and it does not fix the merge in
Fig.~\ref{fig:rs} quantitatively. Star: organizing cusp $(4,2)$.}
\label{fig:phase-diagram}
\end{figure}

Three results arise. First, ``Kuramoto physics'' persists for small $J$. For $J < J_c$, the familiar
async$\to$sync transition is observed (Figure~\ref{fig:rs}, top). For $J>J_c$, however,
new attractors come to life (Figure~\ref{fig:gallery}): a phase-wave state where one coordinate locks
while the other drifts, and a mixed state where both coordinates partially order to different degrees
(Figure~\ref{fig:rs}, bottom). 
 Second, $J>J_c$ induces a nonlinear sync response absent in the Kuramoto model:
as $K$ is increased, $r$ rises in the phase wave state, dips in the mixed state,
then rises again before saturating at $r=1$. Third, the bifurcation
diagram has interesting structure: the transition boundaries of
all four states meet at a single cusp at $(K,J)=(4,2)$ (Figure~\ref{fig:phase-diagram}). 

Now we explain these phenomena analytically. Previous work derived existence
conditions for each state as well as expressions for order parameters \cite{Yoon2022}. Here
we derive stationary neutral conditions and characterize the ordered-state transitions
analytically and numerically.
The results differ in status---exact, approximate, asymptotic, proposed, conjectural, or
numerical---and Table~\ref{tab:status} labels each.

\begin{table}[tbp]
\caption{Status of the main results. \emph{Exact}: closed form or an implicit equation with
elementary terms. \emph{Approximate}: an uncontrolled approximation tested against
simulation. \emph{Asymptotic}: leading terms of an expansion with a stated error.
\emph{Formulated}: condition derived but not solved on the full base state.
\emph{Proposed}: asymptotic form without a controlled derivation. \emph{Conjectural}:
conditional on unproved inputs. \emph{Numerical} and \emph{Interpolation}: computed or
simulation-based.}
\label{tab:status}
\begin{ruledtabular}
\begin{tabular}{lll}
Result & Eq. & Status\\
\hline
Chemotactic reduction & \eqref{eq:ring-x}--\eqref{eq:ring-th} & Asymptotic, $O(\eps_{\rm c})$\\
Async threshold $K_1=4$ & & Exact \cite{Yoon2022}\\
Phase-wave amplitude $r_{\rm pw}$ & & Exact \cite{Yoon2022}\\
Phase-wave neutral $F=0$ & \eqref{eq:F} & Exact$^{d}$\\
\quad slope $\tfrac54$, $c_*$ & \eqref{eq:Jc2} & Exact\\
\quad $-\tfrac{31}{64}$, $\eps^{5/2}$ & \eqref{eq:Jc2} & Numerical\\
Locking diamond & \eqref{eq:diamond} & Exact\\
Sync amplitude $S$ & \eqref{eq:rsync} & Approximate$^{a}$\\
Sync neutral $H=0$ & \eqref{eq:H} & Formulated$^{b}$\\
\quad large-$J$ law $K_3(J)$ & \eqref{eq:Kc3} & Proposed; tested$^{e}$\\
\quad cusp slope $j^*\approx\tfrac{7}{20}$ & \eqref{eq:Jc3} & Conjectural\\
\quad $J_3(K)$ curve, Fig.~\ref{fig:phase-diagram} & & Interpolation\\
Mixed amplitudes $(r,s)$ & \eqref{eq:mixed} & Numerical$^{c}$\\
Mixed-state stability & & Open\\
\end{tabular}
\end{ruledtabular}
\raggedright\footnotesize
$^{a}$$r_{\rm lock}$ as a one-dimensional quadrature of elementary functions; $r_{\rm tongue}$
within center slaving; the opposite-tongue and both-winding populations are neglected;
$S$ from a scalar root.
$^{b}$Stationary ($\lambda=0$) neutral condition; locked term a double quadrature, drift term
numerical; not solved on the full self-consistent base state here.
$^{c}$Full frozen-field orbit average of all oscillators and a two-dimensional root search
(Sec.~IV.C).
$^{d}$Stationary ($\lambda=0$) condition, an implicit equation of elementary terms solved
numerically; its role as the observed transition is supported numerically, and other
unstable branches are not excluded here.
$^{e}$Asymptotic form proposed from a boundary-layer balance; maximum discrepancy about 2.6\% in $J$ over six simulation points.
\end{table}
\subsection{Phase wave} The stability of the async state is $K_1=4$, found
by linearizing around its density $\rho_0 = (2\pi)^{-2}$ \cite{Yoon2022}. The stability of the phase wave is much harder,
because its density is no longer constant. It splits into a locked and a drifting piece:
\begin{equation}
\rho_0 = \frac{\delta(\xi-\xi^*(\nu))}{2\pi}\,\mathbf{1}_{|\nu|<\kappa}
+ \frac{C(\nu)}{2\pi|\nu-\kappa\sin\xi|}\,\mathbf{1}_{|\nu|>\kappa},
\end{equation}
where $\xi^*(\nu)=\arcsin(\nu/\kappa)$ is the stable fixed point,
$\kappa\equiv Kr_{\rm pw}=\sqrt{K(K-4)}$, and $C(\nu)$ is fixed by
normalization. The order parameter $r_{\rm pw}=\sqrt{1-4/K}$ was
derived previously via a self-consistency argument~\cite{Yoon2022}. The expression for $\rho_0$
above is for a $\xi$-wave;
$\eta$-waves exist with equal probability and have the same
stability properties. 

To linearize around this two-piece state, we write $\rho=\rho_0+u\,e^{\lambda t}$ with $u$ small,
substitute into the continuity equation $\partial_t\rho+\partial_\xi(F_\xi\rho)+\partial_\eta(F_\eta\rho)=0$,
and project onto the Fourier modes (Supplemental Material). The result is a linear
operator of form \cite{Mirollo2007}
\begin{equation}
Lu=\lambda u,\qquad L=U+C.\label{eq:Lu}
\end{equation}
The two pieces have a clean meaning. The \emph{uncoupled} part is the linearized
transport along the frozen base flow,
$Uu=-\partial_\xi(F_\xi^0u)-\partial_\eta(F_\eta^0u)$ with $F_\xi^0=\nu-\kappa\sin\xi$,
$F_\eta^0=\mu-Jr_{\rm pw}\sin\xi$. The
\emph{coupled} part is the feedback of the perturbation on the velocities through the
order parameters, $Cu=-\partial_\xi(\rho_0\,\delta F_\xi)-\partial_\eta(\rho_0\,\delta
F_\eta)$, in which $\delta F_\xi,\delta F_\eta$ depend on $u$ only through the two
complex moments $\delta W_\xi=\!\int\! e^{\ii\xi}u$ and
$\delta W_\eta=\!\int\! e^{\ii\eta}u$---so $C$ is finite rank. Inverting the uncoupled
operator, $u=(\lambda-U)^{-1}Cu$, and re-projecting onto those moments collapses
Eq.~\eqref{eq:Lu} to a closed scalar self-consistency
\begin{equation}
\chi(\lambda;K,J)=1,\label{eq:sc}
\end{equation}
the susceptibility $\chi$ being the resolvent matrix element of $(\lambda-U)^{-1}$
sandwiched between the coupling and the moment functionals.

Decomposing $u$ in Fourier harmonics $e^{\ii m\eta}$ of the drifting coordinate
sorts the channels: $m=0$ is exactly the partially locked Kuramoto problem (it
reproduces $K_{1}=4$), the sidebands $|m|\ge2$ are pure transport and stay stable,
and the first genuinely new swarmalator instability is the $m=\pm1$ block. There the
base flow splits the population into $\xi$-\emph{locked} oscillators ($|\nu|<\kappa$,
a discrete pole of the resolvent) and $\xi$-\emph{drifting} oscillators
($|\nu|>\kappa$, the continuous spectrum). The susceptibility is the sum of their
responses, and at neutral stability $\lambda=0$ the self-consistency \eqref{eq:sc}
collapses to the exact, fully elementary equation
\begin{equation}
F(K,J)\equiv\underbrace{\chi_{\ell,1}(0)+\chi_{\ell,2}(0)}_{\text{locked}}
+\underbrace{\chi_d(0)}_{\text{drift}}-1=0,\label{eq:F}
\end{equation}
where $\chi_{\ell,1}$ is the direct $\xi$-response of the locked oscillators,
$\chi_{\ell,2}$ their $\eta$-sideband response, and $\chi_d$ the drift contribution.
Each is a finite sum of arctangents of the frozen orbits; with $\alpha=J/K$ the
locked direct term is, in closed form \cite{Strogatz1991,Mirollo2007},
\begin{equation}
\begin{split}
\chi_{\ell,1}(0)=\frac{K(1-\alpha^2)}{4\pi\alpha}\Big[&(1+\alpha)\arctan\tfrac{(1+\alpha)\kappa}{2}\\
&-(1-\alpha)\arctan\tfrac{(1-\alpha)\kappa}{2}\Big],\label{eq:chil1}
\end{split}
\end{equation}
and $\chi_{\ell,2}(0)$, $\chi_d(0)$ are arctangent sums of the same type. For $\chi_{\ell,2}$
define poles
\begin{align}
q_\pm&=\Big(\tfrac{\sqrt{\kappa^2+4}\pm\kappa}{2}\Big)^2,\\
s_{+,\pm}&=\Big(\tfrac{\sqrt{4+\alpha(\alpha+2)\kappa^2}\pm\kappa(1+\alpha)}{\kappa-2}\Big)^2,\\
s_{-,\pm}&=\Big(\tfrac{\sqrt{4+\alpha(\alpha-2)\kappa^2}\pm\kappa(1-\alpha)}{\kappa-2}\Big)^2,
\end{align}
collected as $(\rho_1,\dots,\rho_6)=(q_+,q_-,s_{+,+},s_{+,-},s_{-,+},s_{-,-})$; then
\begin{align}
\chi_{\ell,2}(0)&=\frac{\kappa J^2}{K\pi(\kappa-2)^4}\sum_{j=1}^{6}
\frac{C_j}{\sqrt{\rho_j}}\arctan\frac{1}{\sqrt{\rho_j}},\label{eq:chil2}\\
C_j&=\frac{N(-\rho_j)}{\prod_{m\ne j}(\rho_m-\rho_j)},\nonumber
\end{align}
with $N(z)=(1-z^2)[L(z)^2M(z)+4\kappa^2zP(z)]$ and auxiliary polynomials
$L(z)=(\kappa+2)+(2-\kappa)z$, $M(z)=(\kappa+4)+(4-\kappa)z$,
$P(z)=(\kappa(1+\alpha^2)+4)+(4-\kappa(1+\alpha^2))z$.
For $\chi_d$ define
\begin{align}
\sigma_{+,\pm}&=\Big(\tfrac{\sqrt{4+\kappa^2(1+2\alpha)}\pm2}{\kappa(1+2\alpha)}\Big)^2,\\
\sigma_{-,\pm}&=\Big(\tfrac{\sqrt{4+\kappa^2(1-2\alpha)}\pm2}{\kappa(1-2\alpha)}\Big)^2,
\end{align}
collected as $(\sigma_1,\dots,\sigma_4)=(\sigma_{+,+},\sigma_{+,-},\sigma_{-,+},\sigma_{-,-})$;
then
\begin{align}
\chi_d(0)&=\frac{64K}{\pi\kappa^3(1+2\alpha)^2(1-2\alpha)^2}\sum_{j=1}^{4}
A_j\,\frac{\arctan\sqrt{\sigma_j}}{\sqrt{\sigma_j}},\label{eq:chid}\\
A_j&=\frac{\sigma_j(\sigma_j+1)}{\prod_{m\ne j}(\sigma_m-\sigma_j)}.\nonumber
\end{align}
Removable cases ($\alpha=0$, $J=K$, $\kappa=2$, $2\alpha=1$) follow by continuity;
roots may be complex and are evaluated by analytic continuation, summing conjugate pairs. Equations~\eqref{eq:F}--\eqref{eq:chid} give an implicit
equation for the phase-wave stationary neutral curve which is exact but opaque. To gain some
insight, we expand it in the small $J$ and large $J$ limits: 
\begin{align}
K_{2}(J)&=4 &&(0\le J\le2),\nonumber\\
J_{2}(K)&=2+\tfrac54\eps-\tfrac{31}{64}\eps^2+O(\eps^{5/2})
 &&(\eps=K-4\to0),\label{eq:Jc2}\\
K_{2}(J)&=J+1+\frac{c_*}{J}+O(J^{-2}) &&(J\to\infty),\nonumber
\end{align}
with $c_*=2+[2+\arctan(1/2)]/\pi$. The slope $\tfrac54$ and $c_*$ are exact; the
$\eps^2$ coefficient $-\tfrac{31}{64}$ and the half-integer $\eps^{5/2}$ term are
high-precision numerical estimates.

These expansions reveal the origin of the cusp $(K,J)=(4,2)$ noted earlier: the
boundary $K_2=4$ is exactly flat for $J\le J_c$ because the locked band is empty
there, leaving only the drift contribution, which is $J$-independent. The mixed
branch leaves the phase wave continuously, $s\sim\sqrt{J_2(K)-J}$; supercriticality is inferred numerically (the cubic coefficient is not computed).

\subsection{Sync}

\subsubsection{Order parameter}

Yoon et al.~\cite{Yoon2022} derived the sync order parameter via the Ott--Antonsen
ansatz, obtaining $S_{\rm OA}=\sqrt{1-4K/(K^2-J^2)}$. This expression is exact on the
decoupled line $J=0$, but for $J\neq0$ it underestimates the order parameter: at
$(K,J)=(6,3)$ the simulated $S=0.462$ is about $39\%$ larger than $S_{\rm OA}=0.333$. At
$J=3$ the formula has no positive real amplitude for $3<K\le2+\sqrt{13}\approx5.61$, although
sync persists down to $K\approx5.5$ (Fig.~\ref{fig:rs}).

The reason lies in the \emph{product} form of the ansatz, not in the OA ansatz as such. At
fixed $(\nu,\mu)$ write $f_{nm}=\avg{e^{-\ii(n\xi+m\eta)}}$. Reference~\cite{Yoon2022} takes a
product of Poisson kernels, $f_{nm}=z_\xi^{\,n}z_\eta^{\,m}$ for $n,m\ge0$ (with conjugates taken
coordinatewise for negative indices): $\xi$ and $\eta$ are treated as statistically independent
for each oscillator, and the sync formula follows from a stationary continuation of this
construction. The $(1,0)$ and $(0,1)$ modes of the exact moment hierarchy give the usual OA
equations for $z_\xi,z_\eta$. The ansatz derivative minus the hierarchy then leaves, in the
first mixed mode, the residual (Supplemental Material, Sec.~S3)
\begin{equation}
R_{11}=\frac{J}{2}\Big[r\,z_\eta(|z_\xi|^2-1)+s\,z_\xi(|z_\eta|^2-1)\Big],
\label{eq:R11}
\end{equation}
with residuals proportional to $J$ in every mixed mode. A stationary product-Poisson density is
therefore generally inconsistent when both the cross-coupling and the mean fields are
nonzero. Special invariant cases remain: $J=0$, incoherence, the phase-wave marginal ($s=0$),
and point measures ($|z_\xi|=|z_\eta|=1$), which describe instantaneous oscillator positions,
moving or not. What the stationary closure cannot represent is the correlated invariant
measure of partially locked oscillators. For an oscillator with $\xi$ locked and $\eta$
winding, $\xi$ librates in step with $\eta$, so the two angles are correlated at fixed
$(\nu,\mu)$. The OA ansatz itself is not at fault: for a single angle whose velocity contains
only its first harmonic, the Poisson family is invariant. The phase-wave marginal has this
structure, and $r_{\rm pw}=\sqrt{1-4/K}$ is exact.

\begin{figure}[t]
\centering
\includegraphics[width=0.95\columnwidth]{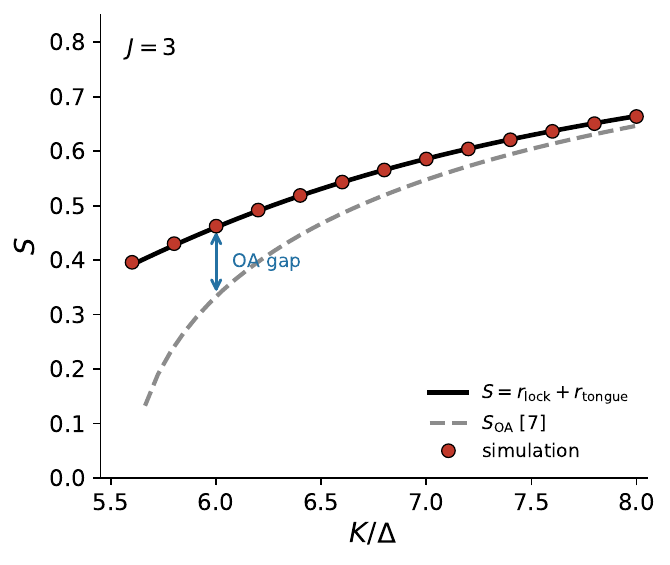}
\caption{Sync order parameter vs $K$ at $J=3$. Solid: locked-plus-tongue approximation
$S\approx r_{\rm lock}+r_{\rm tongue}$ [Eq.~\eqref{eq:rsync}, this work]. Dashed: Ott--Antonsen formula
$S_{\rm OA}$~\cite{Yoon2022}. Markers: direct simulation, the sweep of Fig.~\ref{fig:rs}
($N=2.5\times10^5$; parameters in that caption); only sync points
($|r-s|<0.01$, $K\ge5.6$) are shown. At $K=6$ the simulated $S$ is about $39\%$ above
$S_{\rm OA}$. The OA formula has no positive real amplitude for $3<K\le2+\sqrt{13}\approx5.61$, and
approaches the data deep in sync, where $J/K\to0$.}
\label{fig:SOA}
\end{figure}

The OA discrepancy $S-S_{\rm OA}$ is small in two limits. It vanishes at $J=0$ and, at fixed
$K>4$ away from onset, enters at $O(J^2)$: within the center-slaved approximation below,
writing $S=S_0+c_2J^2+\dots$, the curvatures differ from the OA ones by
$c_2-c_2^{\rm OA}\approx0.0158,\,0.0053,\,0.0014,\,0.0006$ at $K=5,6,8,10$. It also shrinks deep
in sync, $J/K\to0$ (table below and Fig.~\ref{fig:SOA}). It is large near the sync onset for
every $J\neq0$. The discrepancy is not the tongue contribution itself: $\xi$-locked,
$\eta$-drifting oscillators contribute to $S$ even at $J=0$ (about $0.057$ at $K=6$), where they
complete the one-dimensional locked Kuramoto integral; at $J=3$ their center-slaved
contribution falls from $0.108$ at $K=6$ to $0.046$ at $K=8$ as locking becomes more complete.

On the sync branch $r=s=S$ each swarmalator evolves on the frozen torus
\begin{equation}
\dot\xi=\nu-KS\sin\xi-JS\sin\eta,\quad
\dot\eta=\mu-JS\sin\xi-KS\sin\eta,
\label{eq:sync-flow}
\end{equation}
and the order parameter is
\begin{equation}
S=\langle\!\langle\cos\xi\rangle\!\rangle
=\iint h(\nu,\mu)\,\overline{\cos\xi}\,d\nu\,d\mu,
\label{eq:sync-sc}
\end{equation}
with $h(\nu,\mu)=\tfrac{8}{\pi^2[(\nu+\mu)^2+4][(\nu-\mu)^2+4]}$ the bivariate Cauchy
density. The orbit topology of \eqref{eq:sync-flow} partitions the $(\nu,\mu)$ plane into
four classes, each contributing its full orbit average $\overline{\cos\xi}$:
(i)~\emph{both-locked} ($\overline{\cos\xi}=\cos\xi^*$ exactly); (ii)~the \emph{tongue}, $\xi$ bounded
and $\eta$ winding; (iii)~the \emph{opposite tongue}, $\eta$ bounded and $\xi$ winding, whose
$\overline{\cos\xi}$ is generally nonzero because $\dot\xi$ depends on the coupled $\eta$ orbit; and
(iv)~\emph{both-winding}. We evaluate (i) exactly and (ii) in a center-slaved approximation,
$r_{\rm tongue}$, and neglect (iii) and (iv). The result,
$S\approx r_{\rm lock}+r_{\rm tongue}$, is therefore an approximation, which we test against
simulation; its residual combines the slaving error with the omitted classes.

\emph{Both-locked contribution.} At a fixed point $\dot\xi=\dot\eta=0$, Cramer's rule
gives $\sin\xi^*=(K\nu-J\mu)/[S(K^2-J^2)]$, $\sin\eta^*=(K\mu-J\nu)/[S(K^2-J^2)]$.
Writing $a=\sin\xi^*$, $b=\sin\eta^*$ and $A=S(K+J)$, $B=S(K-J)$, the stable branch
($\cos\xi^*,\cos\eta^*>0$) occupies the unit square $|a|,|b|<1$, and the
both-locked contribution reduces to the single arcsine-weighted quadrature
\begin{equation}
r_{\rm lock}(S)=\frac{8}{\pi^2 AB}\int_{-1}^{1}\!\sqrt{1-a^2}\,
[\Phi_a(1)-\Phi_a(-1)]\,da,
\label{eq:rlock}
\end{equation}
where $\Phi_a(b)$ is the elementary primitive of the product of two Lorentzians
centred at $b=\pm a$ with half-widths $\ell=2/A$, $m=2/B$:
\begin{align}
\Phi_a(b)&=\frac{p_a}{2}\ln\frac{(b+a)^2+\ell^2}{(b-a)^2+m^2}
+\frac{q_a}{\ell}\arctan\frac{b+a}{\ell}\nonumber\\
&\quad+\frac{r_a}{m}\arctan\frac{b-a}{m},
\label{eq:Phi}
\end{align}
with $\Delta_a=[4a^2+(\ell+m)^2][4a^2+(\ell-m)^2]$ and
\begin{align*}
p_a&=\tfrac{4a}{\Delta_a},&
q_a&=\tfrac{4a^2-\ell^2+m^2}{\Delta_a},&
r_a&=\tfrac{4a^2+\ell^2-m^2}{\Delta_a}.
\end{align*}
The remaining $a$-integral can be performed in closed form via a semicircle--log
kernel (Supplemental Material, Sec.~S3), yielding a finite sum of complex logarithms that
agrees with direct quadrature to $10^{-10}$.

\emph{Tongue contribution.} Slaving $\xi$ to its center ($\sin\xi^*=a$), the
$\eta$-equation becomes the Adler equation $\dot\eta=M-KS\sin\eta$ with
$M=\mu-JSa$. On the winding branch $|M|>KS$ the orbit average
\begin{equation*}
b\equiv\overline{\sin\eta}=\frac{M-\operatorname{sgn}(M)\sqrt{M^2-(KS)^2}}{KS}
\end{equation*}
inverts to give the tongue chart
\begin{align}
\nu_t(a,b)&=\alpha a+\beta b,\nonumber\\
\mu_t(a,b)&=\beta a+\alpha\tfrac{1+b^2}{2b},\quad
-1<a<1,\ 0<|b|<1,
\label{eq:tongue-chart}
\end{align}
with $\alpha=KS$, $\beta=JS$. The Jacobian is
$D_t=\beta^2+\alpha^2(1-b^2)/(2b^2)$, and the tongue contribution is
\begin{equation}
r_{\rm tongue}(S)=\int_{-1}^{1}\!\sqrt{1-a^2}\,T(a)\,da,
\label{eq:rtongue}
\end{equation}
where the inner $b$-integral $T(a)=\int_{0<|b|<1}h(\nu_t,\mu_t)D_t\,db$ is
elementary: with $P_+(b)=\alpha+2(\alpha+\beta)ab+(\alpha+2\beta)b^2$,
$P_-(b)=-\alpha+2(\alpha-\beta)ab+(2\beta-\alpha)b^2$,
the eight poles of the denominator give
\begin{equation}
T(a)=\operatorname{Re}\sum_{j=1}^{8}\frac{N_a(\rho_j)}{D_a'(\rho_j)}\,
\ln\frac{1-\rho_j}{-1-\rho_j},
\label{eq:T-sumlogs}
\end{equation}
a finite sum of complex logarithms. The outer $a$-integral in \eqref{eq:rtongue} has an
elliptic reduction for generic parameters (Supplemental Material, Sec.~S3); here we evaluate it
by one-dimensional quadrature. At $J=0$, $\beta=0$; the both-locked and
$\xi$-locked/$\eta$-drifting populations then together recover the one-dimensional locked
Kuramoto integral, and $S=r_{\rm lock}+r_{\rm tongue}$ gives $S^2=1-4/K$ exactly.

The approximate self-consistency equation is therefore
\begin{equation}
S=r_{\rm lock}(S;K,J)+r_{\rm tongue}(S;K,J),
\label{eq:rsync}
\end{equation}
which evaluates the locked and partially locked populations separately. Numerical solutions
agree with simulation (Figure~\ref{fig:rs}); representative values are:
\begin{center}
\begin{tabular}{ccccc}
\toprule
$(K,J)$ & $S_{\rm OA}$ & $r_{\rm lock}$ & $r_{\rm tongue}$ & $S$ (sim)\\
\midrule
$(6,3)$ & $0.333$ & $0.352$ & $0.108$ & $0.462$\\
$(7,3)$ & $0.548$ & $0.518$ & $0.068$ & $0.586$\\
$(8,3)$ & $0.647$ & $0.618$ & $0.046$ & $0.664$\\
\bottomrule
\end{tabular}
\end{center}
Simulated values are from the sweep of Fig.~\ref{fig:rs} (caption there).

\subsubsection{Stability and the locking diamond}

The sync state's locking region is not a square but a diamond. In rotated coordinates
$p=(\xi+\eta)/2$, $q=(\xi-\eta)/2$ with $\Omega=(\nu+\mu)/2$, $\Lambda=(\nu-\mu)/2$,
the frozen flow decouples into $\dot p=\Omega-A\sin p\cos q$ and
$\dot q=\Lambda-B\cos p\sin q$. A locked point requires
$u\equiv\Omega/A=\sin p\cos q$ and $v\equiv\Lambda/B=\cos p\sin q$, which satisfy
$|u|+|v|\le1$, i.e.
\begin{equation}
\frac{|\Omega|}{S(K+J)}+\frac{|\Lambda|}{S(K-J)}\le1.
\label{eq:diamond}
\end{equation}
This locking diamond (verified directly in simulation, Supplemental Material
Fig.~S4) replaces the independent-Kuramoto square and is the geometric origin of the
more complex stability structure.

We formulate the exchange-odd stationary neutral condition through the same continuum
linearization as for the phase wave.
The sync state is invariant under the exchange involution
$T:(p,q,\Omega,\Lambda)\mapsto(p,-q,\Omega,-\Lambda)$, which commutes with the
linearized operator $L=U+C$ and splits the spectrum into even and odd sectors.
Since sync$\to$mixed breaks $r=s$, the instability lives in the \emph{odd} sector,
where the finite-rank feedback reduces to a scalar condition
\begin{equation}
H(K,J)\equiv m_a^{\rm lock}(0)+m_a^{\rm drift}(0)-1=0,
\label{eq:H}
\end{equation}
with the locked contribution
\begin{align}
m_a^{\rm lock}(0)&=\frac{S(K^2-J^2)}{2}\iint_{\mathcal{D}}\ell(Au)\,\ell(Bv)\nonumber\\
&\quad\times\Big(\tfrac{1}{c_\xi}+\tfrac{1}{c_\eta}-c_\xi-c_\eta\Big)du\,dv,
\label{eq:malock}
\end{align}
where $\ell(x)=1/[\pi(1+x^2)]$ is the Cauchy density of $\Omega=v'$ and $\Lambda=\omega'$
(half-width 1), $c_\xi=\sqrt{1-(u+v)^2}$,
$c_\eta=\sqrt{1-(u-v)^2}$, and $\mathcal{D}$ is the diamond \eqref{eq:diamond}.
(The prefactor is linear in $S$: differentiating the locked position
$\sin\xi^*\propto1/r$ supplies a $1/r$ that cancels one power of the
change-of-variables Jacobian $\propto S^2$, leaving $S^1$; the $\tfrac12$ is from
exchange symmetrization.)

The drift term $m_a^{\rm drift}$ is a genuinely two-dimensional adjoint problem.
With observable $\psi_a=\sin p\sin q$ and forcing
$f_a=((K+J)\cos p\sin q,\,(K-J)\sin p\cos q)$, the gauge-fixed adjoint
potential $\phi_a$ solves
\begin{align}
&(\Omega-A\sin p\cos q)\,\partial_p\phi_a\nonumber\\
&\quad+(\Lambda-B\cos p\sin q)\,\partial_q\phi_a
=-(\psi_a-\bar\psi_a),
\label{eq:adjoint}
\end{align}
and
\begin{equation*}
m_a^{\rm drift}=\iint_{\mathcal{D}^c}\ell(\Omega)\ell(\Lambda)
\Big(\!\int\rho_0^{\rm drift}f_a\cdot\nabla\phi_a\,dp\,dq\Big)d\Omega\,d\Lambda.
\end{equation*}
Unlike the partially-locked Kuramoto problem---where each drifter moves on a circle
and \eqref{eq:adjoint} is a one-dimensional ODE---here the drift flow is genuinely
two-dimensional with no global first integral, so \eqref{eq:adjoint} is a PDE in two
variables, and we have not found a global closed form for the sync boundary.

Two facts nonetheless tame the boundary. \emph{(i)} At three sampled sync points the two
$\xi$-drifting response channels together contribute about $8$--$11\%$ of the total $m_a$, and
the genuinely both-drifting channel alone about $3$--$5\%$ (Supplemental Material, Fig.~S5),
so the two-dimensional drift is subdominant there. \emph{(ii)} For the $\xi$-locked, $\eta$-winding tongues $\xi$ performs a small
libration driven by the winding $\eta$, governed by the one-dimensional adjoint ODE
\begin{equation}
(M-KS\sin\eta)\,w'(\eta)+\lambda_\xi\,w(\eta)
=\sin\eta-\langle\sin\eta\rangle,
\label{eq:adjoint1d}
\end{equation}
with $\lambda_\xi=KS\cos\xi^*$ and $M=\mu-JS\sin\xi^*$, solvable in closed form
by an integrating factor. The marginal eigenvalue then reduces to a one-dimensional
boundary layer at the $\eta$-locking edge $\dot\eta\to0$, enabling the two
asymptotic descriptions below. A global closed form for $H=0$, valid between these
limits, remains open.

Equation~\eqref{eq:H} is the exchange-odd \emph{stationary} neutral condition, at $\lambda=0$.
Simulations at the tested transitions show a steady, non-oscillatory onset, but we have not
excluded other unstable spectral branches, such as a complex pair or an instability of the
locked sector; as in the partially locked Kuramoto problem, the continuous spectrum need
not be purely imaginary \cite{Mirollo2007}.

\subsubsection{Asymptotics}

In the interior, Fig.~\ref{fig:phase-diagram} uses the simulation-anchored interpolation described in its caption; here we treat $H=0$ in two limits.

\emph{Large $J$.} The $\eta$-locking edge becomes a thin barrier near $\cos q=0$;
rescaling the barrier coordinate $y$, the first-harmonic response obeys
$(2+iy)Z_1=1$, so the in-quadrature kernel is $\mathcal{K}(y)=y/(y^2+4)$, whose
finite part
\begin{equation*}
\mathrm{FP}\!\int_0^\infty\!\Big[\frac{y}{y^2+4}-\frac{\mathbf{1}_{y>1}}{y}\Big]dy=-\log 2
\end{equation*}
sets the $O(1)$ offset. A marginal balance against the $2/\pi$ locked-band
normalization suggests
\begin{equation}
K_3(J)=J+\tfrac{2}{\pi}\log(2J)+\tfrac{4}{\pi}+o(1),
\label{eq:Kc3}
\end{equation}
where, in the proposed balance, the finite part sets the constant. We present
Eq.~\eqref{eq:Kc3} as an asymptotic law proposed from this boundary-layer balance; a matched
derivation with a controlled remainder is not given here. Over the six simulation
points with $J\in[2.6,6.2]$, its maximum discrepancy is about $2.6\%$ in $J$ relative to the
inverse of the proposed curve (Supplemental Material, Fig.~S6).

\emph{Near the cusp.} The marginal condition $m_a(0)=1$ can be expressed through
the Taylor coefficients of the frozen-field map
$\Phi_\xi(r,s)=\langle\!\langle\cos\xi\rangle\!\rangle$ at the sync point $(S,S)$.
Expanding $\Phi_\xi$ to cubic order with the leading-order Cauchy moment expansion gives
the diagonal coefficient $c_x=-K^3/64$ and cross coefficient $c_y=-KJ^2/16$. Subtracting the
fixed-point relation $\Phi_\xi(S,S)=S$, which in the same expansion gives $S^2=\eps/8$ to
leading order in $\eps$, leaves the compact marginal condition
\begin{equation}
(c_x-c_y)+\Delta H\cdot S^2=0,
\label{eq:marginal-map}
\end{equation}
where $\Delta H=H_x(1,1)-H_y(1,1)$ is the exchange-odd directional derivative of
the quintic remainder. Since $c_x-c_y$ vanishes at the cusp with
$\partial_K(c_x-c_y)=-\tfrac12$ and $\partial_J(c_x-c_y)=1$,
\begin{equation}
J_3(K)=2+j^*\eps+O(\eps^2),\quad r_{\rm sync}^2=\tfrac18\eps,
\quad j^*=\tfrac12-\tfrac{\Delta H}{8}.
\label{eq:Jc3}
\end{equation}
The constant $\Delta H$ is set by the double-resonant inner flow at the
cusp. A direct evaluation of its exchange-odd response returns a value consistent
with $\Delta H=6/5$, which would give $j^*=7/20$. The result is therefore conditional on two
unproved inputs, the leading-order moment expansion and $\Delta H=6/5$, and we regard
$j^*\approx7/20$ as a conjecture. It is consistent with simulation but cannot be pinned by it,
since both the wedge width and the symmetry-breaking amplitude vanish as $\eps\to0$, making
a linear fit of $(r-s)^2\to0$ uncontrolled near the cusp.

\subsection{Mixed}
The mixed state is the hardest: its stability is out of scope, and its order
parameters $(r,s)$ have not been computed before. The same locked/tongue/drift
classification applies, now with $r\neq s$ and two tongue populations ($\xi$-locked
with $\eta$ winding, and $\eta$-locked with $\xi$ winding). We obtain $(r,s)$ via
self-consistency.
Freezing $(r,s)$, each swarmalator evolves on the fixed-field torus; the order
parameters must satisfy
\begin{equation}
r=\langle\langle\cos\xi\rangle\rangle,\qquad s=\langle\langle\cos\eta\rangle\rangle,
\label{eq:mixed}
\end{equation}
where $\langle\langle\cdot\rangle\rangle$ denotes the orbit-then-frequency average.
The locked swarmalators contribute in closed form. Setting $\dot\xi=\dot\eta=0$
in Eqs.~\eqref{eq:xi}--\eqref{eq:eta} gives
\begin{equation}
Kr\sin\xi+Js\sin\eta=\nu,\qquad Jr\sin\xi+Ks\sin\eta=\mu,
\label{eq:lock}
\end{equation}
linear in $(\sin\xi,\sin\eta)$ with determinant $rs(K^2-J^2)$. Cramer's rule gives
\begin{equation}
\sin\xi^*=\frac{K\nu-J\mu}{r(K^2-J^2)},\qquad
\sin\eta^*=\frac{K\mu-J\nu}{s(K^2-J^2)},
\label{eq:lockstar}
\end{equation}
on the stable branch $\cos\xi^*,\cos\eta^*>0$, valid inside the locking band
$|\sin\xi^*|,|\sin\eta^*|\le1$. Swarmalators outside this band drift on the frozen
2-torus. The Adler-tongue populations can be treated with the tongue function of the sync
section, and the remainder beyond the locked and center-slaved tongue terms is small
($\lesssim0.3\%$ of $r$, $\lesssim10\%$ of $s$ at sampled points; Supplemental Material). For the curves in Fig.~\ref{fig:rs}, however,
we use no population decomposition: every oscillator is orbit-averaged numerically.

\emph{Numerical method.} We evaluate the right-hand sides of \eqref{eq:mixed} on a
$120\times120$ grid of equal-weight Cauchy quantile nodes for $(v',\omega')$ ($\Delta=1$;
$N=14\,400$ frequency pairs), integrating the frozen flow \eqref{eq:xi}--\eqref{eq:eta} at
fixed $(r,s)$ by RK4 with time step $\Delta t=0.05$: a relaxation window $t_{\rm rel}=200$
followed by an averaging window $t_{\rm avg}=1000$, over which $\cos\xi$ and $\cos\eta$ are
time-averaged with a smooth (Birkhoff) bump weight \cite{Das2017}. Orbits start from
$\xi=\eta=0$; the GPU integration is in single precision, with block sums accumulated in
double precision. This gives the residual
$G(r,s)=\big(\langle\!\langle\cos\xi\rangle\!\rangle-r,\ \langle\!\langle\cos\eta\rangle\!\rangle-s\big)$
on a $17\times21$ grid in $(r,s)$ at each $K$; the mixed root is the crossing of the two zero
contours of $G$ away from the diagonal $r=s$. We locate the root this way rather than by
iteration because the map is nearly degenerate along one direction in the $(r,s)$ plane
(residuals $\sim10^{-3}$ persist over $\Delta s\approx0.1$), so fixed-point or Newton iteration
with a loose tolerance can halt far from the root. We compare the self-consistency at its $11$ computed nodes ($4.95\le K\le5.45$,
$J=3$) with independent long runs at the same $K$ ($N=2.5\times10^5$, relaxation $t=1500$,
averaging $t=500$). The maximum discrepancy is $0.0076$ in $r$ over all nodes and $0.0123$ in $s$
for $K\ge5.1$. The deviation is one-signed---$s$ slightly high and $r$ slightly low at
every node---and grows near the onset of the mixed state, where the self-consistency
overestimates $s$ by up to $0.0333$, comparable to or larger than $s$ itself. There the calculation is not
quantitatively resolved at this frequency resolution. At $K=5$, for example, the saved
residual map gives $G_s\approx3\times10^{-4}$ at the simulated point despite the $0.033$
discrepancy in $s$, so small residuals do not certify the root; the residual grid also starts
at $s=0.02$, above the simulated $s=0.011$ at $K=4.95$. We therefore do not use it to locate the onset or its exponent; the exact endpoint
$K_2$ is available from Eq.~\eqref{eq:F}.

\section{Experimental realization}
The derivation identifies what a realization needs. Tanaka's model already includes
state-dependent chemotactic sensitivity. The realization here additionally requires intrinsic
motility, sensing of chemical contrast rather than absolute level, and first-harmonic spatial
transmission, together with the temporal-response conditions of Sec.~III. Point-source
diffusion does not by itself provide the required spatial filter: for point sources the second
harmonic is never below a quarter of the first. With that caveat, a candidate platform is active Belousov--Zhabotinsky droplets in a single-file
annular microfluidic channel: $x_i$ is angular position, $\theta_i$ the redox phase.
BZ droplets have been coupled in 1D arrays \cite{Delgado2011,Torbensen2017}; active
emulsions self-propel while oscillating \cite{Thutupalli2013}; and droplet speed can
oscillate with the redox cycle \cite{Suematsu2016}. Combining these---mobile droplets
whose aggregation tracks phase similarity---is the outstanding experimental challenge.
Feedback-controlled colloidal swarmalators and self-oscillating Quincke colloids confirm
that tunable phase-position active matter is realistic \cite{Heuthe2025,Leyva2026}.
A proposed measurement is to estimate $J'$ and $K'$ by fitting the two model velocity laws
to tracked positions and phases; the many-body prediction is then the
async$\to$phase-wave$\to$mixed$\to$sync cascade, read from
$R_\pm=|N^{-1}\sum_j e^{\ii(x_j\pm\theta_j)}|$.

\section{Discussion}
Kuramoto's model became central by joining three features: a derivation from a broad
microscopic class, a solvable mean-field theory, and experimental relevance. The
one-dimensional swarmalator now has qualified analogues of all three. It is the
leading-order reduction of a specific class of chemotactic oscillators---intrinsically motile,
with contrast sensing and first-harmonic spatial transmission. It admits exact phase-wave
results and semi-analytical treatments of its ordered states. And BZ droplets on a ring are a
candidate experimental platform.

Several results here are firsts. The phase-wave stationary neutral curve $F(K,J)=0$ is
exact and fully elementary---a finite sum of arctangents---and is consistent with the simulated
transitions.
The product-OA sync amplitude of Ref.~\cite{Yoon2022} misses the correlations of partially
locked oscillators at $J\neq0$ [Eq.~\eqref{eq:R11}]; our orbit-based amplitude
$S\approx r_{\rm lock}+r_{\rm tongue}$ tracks simulation, and the OA discrepancy is $O(J^2)$ at
small $J$ but large near the sync onset (about $39\%$ at $(K,J)=(6,3)$). The locking diamond
\eqref{eq:diamond} is a geometric signature of the two-body coupling absent in
Kuramoto, and the proposed large-$J$ sync boundary $K_3=J+\tfrac{2}{\pi}\log(2J)+\tfrac{4}{\pi}$
is parameter-free, with a maximum discrepancy of about $2.6\%$ in $J$ over the simulated range.

Three open problems stand out. First, the mixed-state stability boundary is
unresolved: the two-dimensional drift that has prevented a closed form for the
sync boundary is even harder to handle there, since $r\neq s$ breaks the exchange symmetry
that reduced the sync problem to a scalar. Second, the cusp slope rests on two unproved inputs: the leading-order moment expansion
(including $S^2\approx\eps/8$) and the inner-flow constant $\Delta H=6/5$, whose
intersection counterterm is identified but not yet evaluated in closed form. Third, sweeps with
Gaussian, Laplace, logistic, and sech disorder show the same qualitative state sequence
(Supplemental Material), but a proof of universality---or a counterexample with
compact-support disorder---is open.

Finally, the chemotactic reduction raises a question about the phase-lagged
generalization. Away from the zero-lag slice the model acquires the lags $\alpha_L$ and
$\beta_L$ of Eqs.~\eqref{eq:ring-x}--\eqref{eq:ring-th}, i.e.\ $\sin(\theta_j-\theta_i+\alpha_L)$
and $\cos(\theta_j-\theta_i-\beta_L)$ kernels. Where ordered states persist, exact
population identities give their mean spatial drift and phase rotation, and simulations find
drifting or rotating continuations for small lags (Supplemental Material, Sec.~S1); their
stability boundaries shift. Whether the cusp survives, splits, or disappears under finite
lag is open.

\end{document}


\title{\texorpdfstring{Supplemental Material:\\ A Solvable Normal Form for Coupled Swarmalators}{Supplemental Material: A Solvable Normal Form for Coupled Swarmalators}}
\author{Kevin P. O'Keeffe}
\affiliation{Starling Research Institute}
\maketitle
\onecolumngrid

This Supplemental Material is organized as follows:
\begin{itemize}
  \item \textbf{S1}: Chemotactic phase--position reduction: explicit realization, approximation statement, general conditions, and numerical verification.
  \item \textbf{S2}: Explicit susceptibility formulas for the phase-wave neutral curve.
  \item \textbf{S3}: Sync order parameter and the limitations of the stationary product-Poisson closure.
  \item \textbf{S4}: Sync stability and its large-$J$ asymptotics.
  \item \textbf{S5}: Mixed-branch self-consistency.
  \item \textbf{S6}: Robustness to the frequency distribution.
\end{itemize} Notation is as in the main text: physical couplings and frequencies
carry primes ($J',K'$, $v_i',\omega_i'$); the transformed couplings are
$(K,J)=\tfrac12(J'+K',\,J'-K')$, with $\kappa=\sqrt{K(K-4)}$ and, in the phase-wave
susceptibilities below, $\alpha=J/K$ (the tongue function reuses $\alpha,\beta,\gamma$ for its
own couplings).
Throughout, the intrinsic disorder is Cauchy: $v_i',\omega_i'$ are independent
Lorentzians of half-width $\Delta=1$, with density $\ell(x)=\pi^{-1}/(1+x^2)$. The
transformed frequencies $\nu=v'+\omega'$, $\mu=v'-\omega'$ are each Lorentzian of
half-width $2$ but are \emph{correlated} (they share $v',\omega'$); their joint density
is
\begin{equation}
h(\nu,\mu)=\frac{8}{\pi^2\big[(\nu+\mu)^2+4\big]\big[(\nu-\mu)^2+4\big]}
=\ell(v')\,\ell(\omega')\,\big|\partial(v',\omega')/\partial(\nu,\mu)\big| .
\end{equation}
Phase-space averages use $\dd\Gamma=\dd\xi\,\dd\eta\,h\,\dd\nu\,\dd\mu$ (or the
corresponding $(p,q,\Omega,\Lambda)$ form), and $\langle\cdot\rangle$ denotes the
frequency average against $h$.

\section{Chemotactic reduction}\label{sec:normalform}
This section derives the 1D swarmalator model as the leading-order phase--position
approximation of a class of Tanaka-type chemotactic oscillators, states the approximation
precisely, gives the general conditions on the oscillator, and checks the reduction
numerically. It supports the main-text statement: \emph{the 1D swarmalator model is the
leading-order phase--position approximation of intrinsically motile Tanaka-type
chemotactic oscillators with contrast sensing and first-harmonic spatial transmission, in
the weak-coupling, zero-lag limit.}

\subsection{What Tanaka's model supplies, and what it does not}\label{sec:tanaka-limits}
Tanaka's chemotactic oscillators~\cite{Tanaka2007} obey
\begin{align}
\dot X_i&=f(X_i)+k\,g\big(S(r_i,t)\big),\label{eq:micro-X}\\
m\ddot r_i&=-\gamma\dot r_i+\beta(X_i)\,\nabla S(r_i,t),\label{eq:micro-R}\\
\tau\partial_tS&=-S+d\nabla^2S+\textstyle\sum_jh(X_j)\,\delta(r-r_j),\label{eq:micro-S}
\end{align}
with internal state $X_i$, position $r_i$, and chemical field $S$. The sensitivity
$\beta(X_i)$ depends on the internal state. This is essential: $\nabla S$ oscillates at
the oscillator frequency, so a constant sensitivity times the oscillating field averages to
zero over a cycle at the order retained here. Constraining motion and transport to a
unit-radius ring and taking the overdamped limit specializes the model directly. Three
features then stand between it and the swarmalator coupling.

\begin{figure}[t]
\centering
\includegraphics[width=0.56\linewidth]{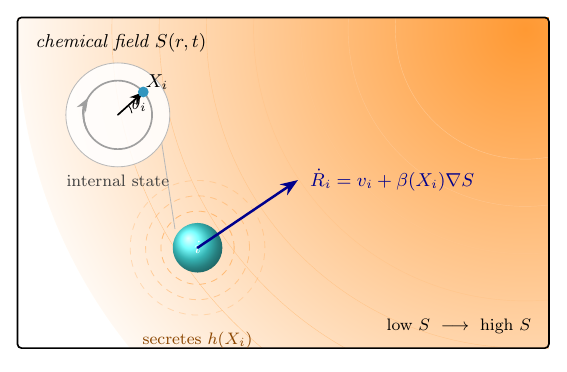}
\caption{Tanaka's chemotactic limit-cycle oscillators~\cite{Tanaka2007}. Each element $i$
carries an internal state $X_i$ on a limit cycle (phase $\theta_i$), secretes a chemical at
the state-dependent rate $h(X_i)$, and moves up the gradient of the field $S(\mathbf r,t)$
with a state-dependent sensitivity; here an intrinsic motility $v_i$ is added. The field in
turn entrains the internal dynamics, closing the phase--position loop.}
\label{fig:chemo}
\end{figure}

\emph{(1) The uniform mode.} The spatially uniform part of $S$ oscillates at the same
frequency as the first spatial mode. Sensed by the oscillators, it adds a leading,
position-independent phase coupling
$\propto N^{-1}\sum_j\Imm[\,q(1+\ii\Omega\tau)^{-1}e^{\ii\delta_{\rm rsp}}e^{\ii(\theta_j-\theta_i)}]$ (with
$\delta_{\rm rsp}$ the rotation of the oscillator's response, zero for the unrotated realization below), which
cannot in general be absorbed into natural frequencies or a rotating frame. With a single
scalar field, it cannot be switched off by tuning the oscillator's response without also
switching off the desired first-harmonic coupling. A separate sensing operation is needed.

\emph{(2) Higher spatial harmonics.} For point sources on the ring the non-constant cosine
coefficients of the chemical response are $g_n\propto(1+dn^2+\ii\Omega\tau)^{-1}$, so
\begin{equation}
\Big|\frac{g_2}{g_1}\Big|^2=\frac{(1+d)^2+(\Omega\tau)^2}{(1+4d)^2+(\Omega\tau)^2}\ \ge\ \frac1{16},
\label{eq:g2g1}
\end{equation}
since $16[(1+d)^2+(\Omega\tau)^2]-[(1+4d)^2+(\Omega\tau)^2]=15+24d+15(\Omega\tau)^2>0$. In the
gradient the second harmonic carries relative weight $2|g_2/g_1|\ge\tfrac12$. Neither the
ring geometry nor fast diffusion makes higher harmonics parametrically small. First-harmonic
transmission must therefore be imposed, either as a declared truncation or through a
spatial filter on secretion or sensing. For example, a wrapped-Gaussian source of width
$\sigma$ suppresses mode $n$ relative to mode $1$ by $e^{-(n^2-1)\sigma^2/2}$.

\emph{(3) Intrinsic motility.} Tanaka's elements move only by chemotaxis. The spatial
natural frequencies $v_i'$ of the nonidentical model must be added as an intrinsic
tangential drive.

\subsection{An explicit realization}
We make the three additions explicitly. Take isochronous Stuart--Landau oscillators with
internal coordinates $(u_i,w_i)$, $\Omega_i=\Omega+\omega_i'$, and
\begin{align}
\dot u_i&=\lambda(1-u_i^2-w_i^2)u_i-\Omega_iw_i+a\,C(x_i,t),\nonumber\\
\dot w_i&=\Omega_iu_i+\lambda(1-u_i^2-w_i^2)w_i,\nonumber\\
\dot x_i&=v_i'+b\,u_i\,\partial_xS(x_i,t),\label{eq:explicit}\\
\tau\partial_tS&=-S+d\,\partial_x^2S+\frac qN\sum_ju_j\big[1+\cos(x-x_j)\big],\nonumber\\
C&=S-\bar S,\qquad \bar S(t)=\frac1{2\pi}\int_0^{2\pi}S(x,t)\,dx.\nonumber
\end{align}
Here $a$ is the linearized chemical response of the oscillator (absorbing Tanaka's $k\,g'(0)$),
$b\,u_i=\beta(X_i)/\gamma$ is the state-dependent sensitivity divided by friction, $q\,u_j$
is the secretion (linear in the state, with zero cycle mean), and the non-negative profile
$1+\cos(x-x_j)$ spreads it over the ring. The $1/N$ is a mean-field normalization of the
source strength. Contrast sensing $C=S-\bar S$ acts on the oscillator's input; the field
keeps its ordinary decay $-S$, so every mode is damped.

In polar form, $u_i=\rho_i\cos\phi_i$, $w_i=\rho_i\sin\phi_i$,
\begin{equation}
\dot\rho_i=\lambda(1-\rho_i^2)\rho_i+aC\cos\phi_i,\qquad
\dot\phi_i=\Omega_i-\frac{aC}{\rho_i}\sin\phi_i,
\end{equation}
so radial perturbations relax at rate $2\lambda$, the phase response is $-a\sin\phi$, the
sensitivity is $b\cos\phi$, and the secretion is $q\cos\phi$. The oscillator is isochronous,
so there is no shear correction.

The source contains only the uniform and first spatial modes. For initial fields in the span of
$1$, $\cos x$ and $\sin x$ the field stays exactly in the three-mode subspace
$S=S_0(t)+S_c(t)\cos x+S_s(t)\sin x$ (other initial modes $n\ge2$ decay as
$e^{-(1+dn^2)t/\tau}$), with
\begin{align}
\tau\dot S_0&=-S_0+\tfrac qN\textstyle\sum_ju_j,\nonumber\\
\tau\dot S_c&=-(1+d)S_c+\tfrac qN\textstyle\sum_ju_j\cos x_j,\\
\tau\dot S_s&=-(1+d)S_s+\tfrac qN\textstyle\sum_ju_j\sin x_j.\nonumber
\end{align}
Contrast sensing removes $S_0$ from the oscillators exactly, and $S_0$ has zero gradient.
Within this subspace no spatial truncation is involved.

\subsection{Cycle averaging}
Let $\eps_{\rm c}$ be the coupling scale. Write $a=\eps_{\rm c}a_0$, $b=\eps_{\rm c}b_0$,
$v_i'=\eps_{\rm c}V_i$ and $\omega_i'=\eps_{\rm c}W_i$, with $a_0,b_0,q,V_i,W_i$ bounded independently of
$\eps_{\rm c}$. The field $S$ is the scaled field; equivalently, the physical chemical perturbation
is $\eps_{\rm c}S$, with weak secretion $\eps_{\rm c}q$ and unscaled responses $a_0,b_0$. Take $q>0$
without loss of generality, absorbing its sign into $a_0$ and $b_0$. Then $\rho_i=1+O(\eps_{\rm c})$, and positions and phase
offsets $\theta_i=\phi_i-\Omega t$ are slow. The first spatial mode responds to secretion at
frequency $\Omega$ with complex gain
\begin{equation}
H=\frac{q}{1+d+\ii\Omega\tau}=\frac{q}{L}\,e^{-\ii\delta},\qquad
L=\sqrt{(1+d)^2+(\Omega\tau)^2},\quad \delta=\arctan\frac{\Omega\tau}{1+d}.
\end{equation}
Holding $x_i,\theta_i$ fixed over a cycle and using
$\langle\cos\phi_i\cos\phi_j\rangle=\tfrac12\cos(\theta_j-\theta_i)$ and
$\langle-\sin\phi_i\cos\phi_j\rangle=\tfrac12\sin(\theta_j-\theta_i)$ (checked symbolically),
\begin{align}
\dot x_i&=v_i'+\frac b{2N}\sum_j\Ree\!\big[H\,e^{\ii(\theta_j-\theta_i)}\big]\sin(x_j-x_i),\nonumber\\
\dot\theta_i&=\omega_i'+\frac a{2N}\sum_j\Imm\!\big[H\,e^{\ii(\theta_j-\theta_i)}\big]\cos(x_j-x_i).
\label{eq:lagged-avg}
\end{align}
The self terms $j=i$ are kept, as in the target model; at finite lag they give a phase
drift that must not be dropped. Writing $H=(q/L)e^{-\ii\delta}$ gives the phase-lagged
swarmalator of the main text, with $J'=bq/(2L)$, $K'=aq/(2L)$ and lags
$\beta_L=-\alpha_L=\delta$. In this realization the two
lags are tied to the one chemical lag; independent lags need more general response
waveforms (Sec.~\ref{sec:general-resp}).

\emph{Zero lag by fast chemistry.} For $\Omega\tau/(1+d)\ll1$, $H\to q/(1+d)$ is real and
\begin{equation}
J'=\frac{bq}{2(1+d)},\qquad K'=\frac{aq}{2(1+d)}.
\end{equation}
The signs of $a$ and $b$ are free, so $J'$ and $K'$ can take either sign.

\emph{Zero lag by compensation.} Alternatively, keep $\tau$ finite and rotate both the
oscillator's chemical input and its sensitivity by $\delta$: replace $a\,C$ in the $u$
equation by $a\,C(\cos\delta,\sin\delta)$ in $(u,w)$, and $b\,u_i$ by
$b(u_i\cos\delta+w_i\sin\delta)$. The effective gain becomes $He^{\ii\delta}=q/L$, which
is real, and
\begin{equation}
J'=\frac{bq}{2L},\qquad K'=\frac{aq}{2L}.
\end{equation}
This is a physical choice of response direction relative to secretion, not a change of
phase coordinates; rephasing every oscillator cannot remove a lag. It uses the common base
frequency $\Omega$, and detunings change the transfer only at next order.

\subsection{Approximation statement}\label{sec:approx}
Fix a finite $N$, a time horizon $T$, and $\lambda,\Omega,d>0$, $0<\tau<\infty$, with
$a_0,b_0,q,V_i,W_i$ bounded independently of $\eps_{\rm c}$. Start the oscillators on the cycle, $\rho_i=1$, and the
three field modes on their periodic response to the initial positions and phases. Let
$z_{\rm micro}=(x,\phi-\Omega t)$ be the microscopic solution (continuous angular lifts) and
$z_{\rm avg}$ the solution of the averaged equations~\eqref{eq:lagged-avg} with the same
initial data. Then, for sufficiently small $\eps_{\rm c}$, there is a constant $C_T$ independent of
$\eps_{\rm c}$ with
\begin{equation}
\sup_{0\le t\le T/\eps_{\rm c}}\ \big|z_{\rm micro}(t)-z_{\rm avg}(t)\big|\le C_T\,\eps_{\rm c}.
\end{equation}
With compensated responses $z_{\rm avg}$ is the zero-lag swarmalator model. With
uncompensated responses the zero-lag model is approached with error
$O(\eps_{\rm c}+\Omega\tau/(1+d))$, since
$|H-q/(1+d)|=\frac{|q|}{1+d}\,\frac{\delta_c}{\sqrt{1+\delta_c^2}}$ with $\delta_c=\Omega\tau/(1+d)$;
taking $\tau\propto\eps_{\rm c}$ restores $O(\eps_{\rm c})$.

\emph{Proof sketch.} (i) \emph{Radii.} The radial equation is a stable relaxation forced at
$O(\eps_{\rm c})$; the field modes obey stable linear equations forced by bounded sources. A
bootstrap in a fixed tube around $\rho=1$ gives $\rho_i-1=O(\eps_{\rm c})$ on $[0,T/\eps_{\rm c}]$, so phases
stay regular. (ii) \emph{Field.} Let $F_\star(t;x,\theta)$ be the prepared periodic field
response at the current positions and phases. The error $E=F-F_\star$ obeys a stable linear
equation forced at $O(\eps_{\rm c})$, both by amplitude errors and by the slow motion of the sources;
variation of constants gives $E=O(\eps_{\rm c})$. (iii) \emph{Slow equation.} Substituting (i)--(ii)
gives $\dot z=\eps_{\rm c}\,\mathcal F(\Omega t,z)+O(\eps_{\rm c}^2)$ with $\mathcal F$ smooth and
$2\pi/\Omega$-periodic. (iv) \emph{Averaging.} A bounded periodic corrector
$z=Z+\eps_{\rm c} U(\Omega t,Z)$ removes the oscillating $O(\eps_{\rm c})$ term, leaving
$\dot Z=\eps_{\rm c}\bar{\mathcal F}(Z)+O(\eps_{\rm c}^2)$; Gronwall's inequality on $t\le T/\eps_{\rm c}$ gives the
bound~\cite{SVM}. The radial and field estimates (i)--(ii) are the problem-specific input;
the averaging step is standard.

\emph{Scope.} This is a finite-time statement on the interaction time scale for a finite
population. It is not uniform as $T\to\infty$, near bifurcations, or as $N\to\infty$ with
unbounded (Cauchy) disorder; the thermodynamic-limit theory of the main text treats the
averaged model as given. Bounded initial field mismatch and radii in a compact annulus
inside the basin add integrable transients whose effect on the slow variables is $O(\eps_{\rm c})$.
Finite inertia, $t_v=m/\gamma$ fixed, also leaves the averaged equations unchanged: the
periodic velocity response has the same cycle mean as its forcing, and its oscillating part
moves positions only at $O(\eps_{\rm c})$. Inertia on the interaction time scale, $t_v=O(1/\eps_{\rm c})$,
would need a different reduced system.

\subsection{General limit cycles}\label{sec:general-resp}
For a hyperbolically attracting limit cycle with asymptotic phase $\phi$, let $h(\phi)$ be
the secretion waveform, and $B(\phi)$ and $Z(\phi)$ the \emph{unscaled} tangential sensitivity
(divided by friction) and phase response to the chemical input, all evaluated on the cycle;
their physical prefactors are $\eps_{\rm c}B$ and $\eps_{\rm c}Z$. Expand
$h=\sum_mh_me^{\ii m\phi}$, and similarly $B$ and $Z$. With first-harmonic contrast
transmission, $T_m=(1+d+\ii m\Omega\tau)^{-1}$, the leading averaged equations on slow time
$s=\eps_{\rm c} t$ are
\begin{align}
\frac{dx_i}{ds}&=V_i+\frac1N\sum_jF_x(\theta_j-\theta_i)\sin(x_j-x_i),\\
\frac{d\theta_i}{ds}&=W_i+\frac1N\sum_jF_\theta(\theta_j-\theta_i)\cos(x_j-x_i),
\end{align}
with $F_x(\Delta)=\sum_mB_{-m}h_mT_me^{\ii m\Delta}$ and $F_\theta(\Delta)=\sum_mZ_{-m}h_mT_me^{\ii m\Delta}$,
both real. These reduce to the zero-lag swarmalator model exactly when
\begin{gather}
B_0h_0=Z_0h_0=0,\qquad B_{-m}h_m=Z_{-m}h_m=0\ \ (m\ge2),\nonumber\\
B_{-1}h_1T_1=\tfrac12J_0\in\mathbb R,\qquad Z_{-1}h_1T_1=-\tfrac\ii2K_0,
\end{gather}
with $J'=\eps_{\rm c} J_0$, $K'=\eps_{\rm c} K_0$. The conditions constrain \emph{correlations}, not each
waveform: purely sinusoidal $B$ and $Z$ reject all higher harmonics of the secretion, and
purely sinusoidal secretion allows any higher harmonics in $B$ and $Z$. For the explicit
realization, $B_{-1}=\tfrac{b_0}2e^{\ii\delta}$, $Z_{-1}=-\tfrac{\ii a_0}2e^{\ii\delta}$ (compensated)
and $h_1=\tfrac q2$, which gives $J_0=b_0q/(2L)$ and $K_0=a_0q/(2L)$, so $J'=\eps_{\rm c}J_0$ and
$K'=\eps_{\rm c}K_0$ agree with the explicit construction ($V_i$, $W_i$ are the unscaled drives and
detunings).

\emph{Mean sensitivity and mean secretion.} The condition $B_0h_0=0$ matters. Constant
secretion $h_0$ with constant sensitivity $B_0$ produces a quasi-static first-harmonic
field and the phase-independent attraction
\begin{equation}
\dot x_i\ni\eps_{\rm c}\,\frac{B_0h_0}{1+d}\,\frac1N\sum_j\sin(x_j-x_i),
\end{equation}
at the same order as $J'$. This term is present even though $h_0$ is state-independent,
because unevenly placed constant sources still create a gradient; contrast sensing does not
remove it. It vanishes if $B_0=0$ or $h_0=0$. For a general oscillator, linear secretion in
$X$ does not by itself guarantee $h_0=0$, since the cycle can have a non-zero mean.

\emph{Second-order drift.} ``Constant sensitivity gives no drift'' holds only at the order
retained. A traveling oscillatory velocity $\dot x=A\cos(x-\Omega t)$ has mean drift
$\Omega-\sqrt{\Omega^2-A^2}=A^2/(2\Omega)+\dots$. With weak secretion, $A=O(\eps_{\rm c})$, this is
beyond the reduction; in a near-Hopf ordering with $A=O(\mu^{1/2})$ it is $O(\mu)$ and can
compete with retained terms, so the ordering must be checked before importing a general Hopf
reduction.

\emph{Shear.} For the Stuart--Landau oscillator $\dot z=(1+\ii\omega_0)z-(1+\ii c)|z|^2z+p$
with $z=\rho e^{\ii\psi}$, the isochron phase $\vartheta=\psi-c\ln\rho$ obeys
$\dot\vartheta=\omega_0-c+\Imm[(1-\ii c)\,p\,e^{-\ii\psi}]/\rho$. The effective phase
response is therefore rotated by $(1-\ii c)$; using the geometric phase without this factor
gives the wrong response correlation $Z_{-1}$.

\emph{Weak secretion as the small parameter.} Writing the physical field as $\eps_{\rm c} S$, a smooth
chemical response with $g(0)=0$ satisfies $k\,g(\eps_{\rm c} C)=\eps_{\rm c}\,k\,g'(0)\,C+O(\eps_{\rm c}^2)$. This
makes the linearization controlled. A small prefactor $k$ multiplying an $O(1)$ field would
not: a nonlinear $g$ then generates many-body terms at the coupling order. The reduction
therefore does not require proximity to a Hopf bifurcation, only a stable limit cycle and
weak secretion; close to Hopf the amplitude relaxation is itself slow and the ordering must
be adjusted.

\subsection{Numerical verification}
We integrated the explicit parent system~\eqref{eq:explicit}---Cartesian oscillator
coordinates, positions, and all forced field modes, retaining amplitude changes and chemical
memory---and compared it with the averaged equations~\eqref{eq:lagged-avg}, implemented
independently as pairwise phase--position equations on slow time. Both start from the same
positions and phases; nothing is fitted and no reference frame is aligned. The error is the
maximum, over $0\le\eps_{\rm c} t\le6$, of the RMS wrapped-angle difference across all positions and
phase offsets, sampled at $\ge24$ points per fast cycle. Parameters: $\Omega=2$, $\lambda=1.4$,
$q=1.2$, $d=0.7$. Unless indicated otherwise, $N=8$, seed $11$, $a_0=1.1$ and $b_0=1.4$; the
joint-limit cases use $a_0=-1.1$; the two-particle cases use $a_0=0.9$, $b_0=-1.2$, $\tau=0$
and seed $15$; the negative-motion and both-negative cases use the indicated sign reversals
(seeds $29$ and $7$). Unscaled drives and detunings are drawn uniformly on $[-0.12,0.12]$ and
centered. Finite chemistry uses $\tau=0.8$, an uncompensated chemical lag of about $43^\circ$. Integration
used DOP853 with ${\rm rtol}=2\times10^{-10}$, ${\rm atol}=2\times10^{-12}$; three cases repeated at
$100\times$ tighter tolerance (one with the implicit Radau method) changed the maximum RMS
angle by at most $3.8\times10^{-9}$\,rad (maximum single-coordinate change $5.4\times10^{-9}$\,rad). In all, 42 trajectory cases were run.

\begin{table}[h]
\caption{Microscopic vs averaged dynamics: error (rad) at the largest and smallest $\eps_{\rm c}$, and
the observed order (log--log slope over the three smallest $\eps_{\rm c}$; two for the sign checks).}
\label{tab:chemo-conv}
\begin{ruledtabular}
\begin{tabular}{lcccc}
Case & $\eps_{\rm c}$ range & largest $\eps_{\rm c}$ & smallest $\eps_{\rm c}$ & order\\
\hline
Instantaneous chemistry & 0.16--0.01 & 0.0254 & 0.00159 & 1.00\\
Fast chemistry, $\tau=\eps_{\rm c}$ (zero-lag target) & 0.16--0.01 & 0.142 & 0.00947 & 0.99\\
Finite chemistry, lagged model & 0.16--0.01 & 0.00908 & 0.000565 & 1.00\\
Finite chemistry, compensated & 0.16--0.01 & 0.0177 & 0.00114 & 0.99\\
Two particles & 0.08--0.01 & 0.00709 & 0.000895 & 0.99\\
Finite inertia, $m/\gamma=0.7$ & 0.04--0.01 & 0.00678 & 0.00174 & 0.98\\
Off-cycle radii, zero initial field & 0.04--0.01 & 0.00532 & 0.00135 & 0.99\\
$J'<0$, $N=12$ & 0.02--0.01 & 0.000482 & 0.000241 & 1.00\\
$J',K'<0$, $N=4$ & 0.02--0.01 & 0.000718 & 0.000359 & 1.00\\
\end{tabular}
\end{ruledtabular}
\end{table}

\begin{figure}[h]
\centering
\includegraphics[width=\linewidth]{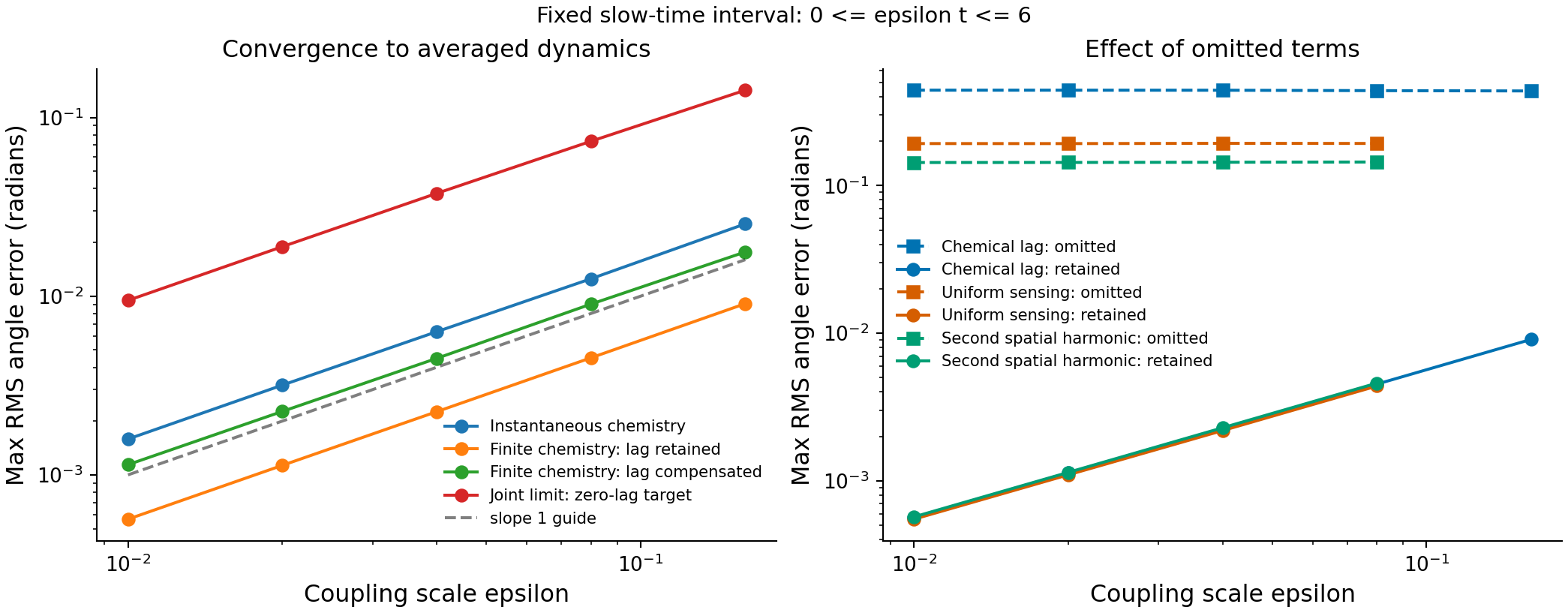}
\caption{Numerical check of the chemotactic reduction. \emph{Left}: maximum RMS angle error
between the explicit microscopic model~\eqref{eq:explicit} and its averaged equations over
$0\le\eps_{\rm c} t\le6$, vs coupling scale $\eps_{\rm c}$, for instantaneous chemistry, finite chemistry with
the lag retained, finite chemistry with compensated responses, and the joint limit
$\tau=\eps_{\rm c}$ compared with the zero-lag target; dashed: slope 1. \emph{Right}: at fixed
$\eps_{\rm c}$, models omitting the chemical lag, the uniform-mode phase coupling, or the second
spatial harmonic (squares, dashed) fail to converge; complete models (circles) converge at
first order.}
\label{fig:chemo-conv}
\end{figure}

Every case converges at first order in $\eps_{\rm c}$ (Table~\ref{tab:chemo-conv},
Fig.~\ref{fig:chemo-conv}). The controls in Fig.~\ref{fig:chemo-conv} (right) test the three
features of Sec.~\ref{sec:tanaka-limits}. At $\eps_{\rm c}=0.01$, averaged models that \emph{omit} the chemical lag, the
uniform-mode phase coupling (when the oscillators sense $S$ rather than $C$), or a second
spatial harmonic (source $1+\cos x+0.6\cos2x$) have errors $0.443$, $0.193$ and $0.143$\,rad
that do not decrease with $\eps_{\rm c}$; the corresponding complete models converge at first order
with errors $\approx5.5\times10^{-4}$. An independent two-particle cycle quadrature agrees with
the averaged velocities to $2\times10^{-16}$. These tests concern the explicit realization; they
do not test the unfiltered point-source model or infinite-time phase-diagram equivalence.

\subsection{Robustness off the zero-lag slice}
Averaging the phase-lagged equations above over the population gives two \emph{exact}
identities for the mean position $\bar x=N^{-1}\sum_i x_i$ and mean phase
$\bar\theta=N^{-1}\sum_i\theta_i$,
\begin{equation}
\dot{\bar x}=\bar v'+\frac{J'\sin\beta_L}{2}\,(s^2-r^2),\qquad
\dot{\bar\theta}=\bar\omega'+\frac{K'\sin\alpha_L}{2}\,(r^2+s^2),
\label{eq:drift-rotation}
\end{equation}
so a motility lag ($\beta_L\neq0$) drives a collective spatial \emph{drift} whenever
$r\neq s$, and a phase-response lag ($\alpha_L\neq0$) a collective \emph{rotation}
whenever $r^2+s^2>0$. Where the ordered states persist, these identities give their mean
spatial and phase rates; the simulations below find drifting or rotating continuations for
the tested small lags. Async ($r=s=0$) does neither; sync ($r=s$)
rotates but does not drift; the phase wave ($r\neq s$) becomes a traveling wave.

Deterministic-Cauchy simulations confirm Eq.~\eqref{eq:drift-rotation} to machine
precision (RMS residuals $E_x\le1.9\times10^{-13}$, $E_\theta\le9.9\times10^{-14}$)
and show a clean separation of the two lags: $\alpha_L$-only preserves all four state
labels and adds only phase rotation, while $\beta_L$-only drives drift and is what
eventually destabilizes the phase wave. Across a small-lag grid the async, mixed, and
sync states persist; the phase wave is the fragile one. At a representative
nonidentical phase-wave point $(J',K')=(8,-2)$ the traveling wave is longitudinally
robust up to $\beta_L\approx0.34$, with an intermittent, path-dependent finite-$N$
state-label breakdown near $\beta_L\approx0.36$--$0.42$ ($N=2000$), not a clean
square-root onset. The corresponding identical-frequency wave's comoving-frame
Jacobian crosses only near $\beta_L\approx0.616$, so the heterogeneous breakdown is not
naive $J'\cos\beta_L$ weakening but a \emph{transverse} fragility: motility lag plus
frequency disorder excites the rainbow mode left out of the zero-lag slice. These finite-$N$ tests
find continuations of the sampled states for small lags; a complete off-slice taxonomy and an
experimental realization remain open.

\section{Phase-wave stability}

\subsection{Continuum linearization and the \texorpdfstring{$m=\pm1$}{m=+-1} channel}
In the transformed variables the continuum characteristics are
\begin{align}
\dot\xi&=\nu+K\,\Imm\!\big(W_\xi e^{-\ii\xi}\big)+J\,\Imm\!\big(W_\eta e^{-\ii\eta}\big),\\
\dot\eta&=\mu+J\,\Imm\!\big(W_\xi e^{-\ii\xi}\big)+K\,\Imm\!\big(W_\eta e^{-\ii\eta}\big),
\end{align}
with $W_\xi=\int e^{\ii\xi}\rho\,\dd\Gamma$, $W_\eta=\int e^{\ii\eta}\rho\,\dd\Gamma$,
$r=|W_\xi|$, $s=|W_\eta|$, and frequencies $\nu=v'+\omega'$, $\mu=v'-\omega'$. (With the
phases fixed to zero, $K\,\Imm(W_\xi e^{-\ii\xi})=-Kr\sin\xi$, recovering the main-text
form.) The phase wave is the branch $(r,s)=(r_{\rm pw},0)$ (the $\xi$-locked
representative; the $\eta$-locked partner $(0,r_{\rm pw})$ is equivalent under the
$\xi\leftrightarrow\eta$ exchange symmetry) with
\begin{equation}
r_{\rm pw}(K)=\sqrt{1-4/K}\quad(K>4),\qquad K_{1}=4 ,
\end{equation}
along which $\xi$ is partially locked---base flow $\dot\xi=\nu-\kappa\sin\xi$,
$\kappa\equiv Kr_{\rm pw}=\sqrt{K(K-4)}$---while $\eta$ drifts; this is why the problem
stays close to the partially locked Kuramoto calculation of Mirollo and
Strogatz~\cite{Strogatz1991,Mirollo2007}.

Here $\rho_0$ is the phase-wave stationary density. With $\eta$ uniform ($s=0$) it
factorizes as $\rho_0(\xi,\eta\mid\nu,\mu)=\varrho(\xi\mid\nu)/(2\pi)$, where the
$\xi$-marginal is a locked $\delta$-spike or a drift density $\propto1/|\dot\xi|$,
\begin{equation}
\varrho(\xi\mid\nu)=
\begin{cases}
\delta(\xi-\xi^*),\quad \sin\xi^*=\nu/\kappa,\ \cos\xi^*>0, & |\nu|<\kappa\ (\text{locked}),\\[6pt]
\dfrac{\sqrt{\nu^2-\kappa^2}}{2\pi\,|\nu-\kappa\sin\xi|}, & |\nu|>\kappa\ (\text{drifting}),
\end{cases}
\end{equation}
the drift branch being the invariant density of $\dot\xi=\nu-\kappa\sin\xi$ (normalized
to $\int_0^{2\pi}\varrho\,\dd\xi=1$). Writing $\rho=\rho_0+u\,e^{\lambda t}$ in the
continuity equation
$\partial_t\rho+\partial_\xi(F_\xi\rho)+\partial_\eta(F_\eta\rho)=0$ and keeping first
order gives the eigenvalue problem $Lu=\lambda u$ with $L=U+C$. The \emph{uncoupled}
part is transport along the frozen base flow,
\begin{equation}
Uu=-\partial_\xi(F_\xi^0u)-\partial_\eta(F_\eta^0u),\qquad
F_\xi^0=\nu-\kappa\sin\xi,\quad F_\eta^0=\mu-Jr_{\rm pw}\sin\xi,
\end{equation}
and the \emph{coupled} part is the finite-rank feedback of the perturbation on the
velocities through the two moments $\delta W_\xi=\int e^{\ii\xi}u\,\dd\Gamma$,
$\delta W_\eta=\int e^{\ii\eta}u\,\dd\Gamma$,
\begin{equation}
Cu=-\partial_\xi(\rho_0\,\delta F_\xi)-\partial_\eta(\rho_0\,\delta F_\eta),\qquad
\begin{aligned}
\delta F_\xi&=K\,\Imm(\delta W_\xi e^{-\ii\xi})+J\,\Imm(\delta W_\eta e^{-\ii\eta}),\\
\delta F_\eta&=J\,\Imm(\delta W_\xi e^{-\ii\xi})+K\,\Imm(\delta W_\eta e^{-\ii\eta}).
\end{aligned}
\end{equation}
Decomposing $u$ in Fourier modes
$e^{\ii m\eta}$ of the drifting coordinate sorts the channels: $m=0$ is exactly the
partially locked Kuramoto problem (it reproduces $K_{1}=4$), the sidebands $|m|\ge2$
are pure transport and stay stable, and the first genuinely new swarmalator
instability is the $m=\pm1$ block. Inverting $U$ and reprojecting onto the moments
collapses that block to a scalar self-consistency $\chi(\lambda;K,J)=1$; at neutral
stability $\lambda=0$,
\begin{equation}
F(K,J)=\chi_{\ell,1}(0)+\chi_{\ell,2}(0)+\chi_d(0)-1=0 ,
\end{equation}
where $\chi_{\ell,1}$ is the direct $\xi$-response of the locked oscillators,
$\chi_{\ell,2}$ their $\eta$-sideband response, and $\chi_d$ the drift contribution.
Each is an elementary integral; equivalently a finite arctangent sum.

\begin{figure}[t]
\centering
\includegraphics[width=0.62\linewidth]{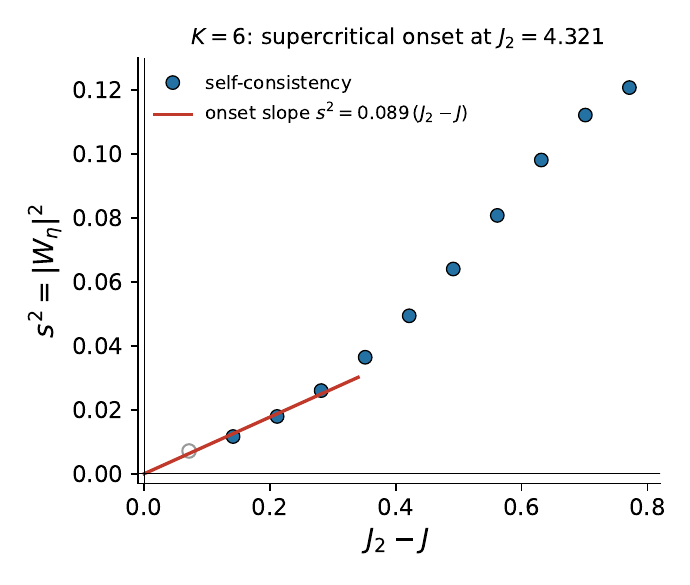}
\caption{Supercritical onset of the mixed state at the phase-wave boundary. Mixed-branch
self-consistency at $K=6$ (smaller order parameter $s=|W_\eta|$): $s^2$ rises continuously
and linearly from zero as $J\to J_2(K)$ (exact $J_2(6)=4.321$), with positive onset slope
(red line, fit over the onset region $J_2-J<0.3$), supporting a supercritical pitchfork
($g_3>0$, no hysteresis). The hollow point is affected by the finite-grid $s$-floor;
the super-linear rise away from onset is the higher-order branch shape.}
\label{fig:verif-onset}
\end{figure}

We solve $\chi=1$ at the \emph{real} point $\lambda=0$, the stationary neutral condition. We do
not prove that this exhausts the instabilities: as in the partially locked Kuramoto state,
the continuous spectrum includes a contracting locked part as well as the drifting part on
$i\mathbb R$ \cite{Mirollo2007}, and a complex pair crossing elsewhere is not excluded
analytically. The evidence that the loss of stability is steady is numerical. A frequency-domain winding (Nyquist) test
of $\chi(\lambda)$ over $\Ree\lambda\ge0$ is delicate here precisely because the continuous
spectrum lies on $i\mathbb R$ itself, so the test contour runs alongside it; we therefore verify
the onset \emph{character} directly in the time domain. At both the phase-wave$\to$mixed and
sync$\to$mixed boundaries the symmetry-breaking amplitude grows \emph{monotonically} from zero
with no oscillatory transient and no limit cycle at the sampled parameters
(Fig.~\ref{fig:verif-onset} and the cusp-slope sweeps). These time-domain transitions are
consistent with a continuous steady bifurcation; this numerical evidence does not exclude
oscillatory or other instabilities at untested parameters.

\paragraph{Integral form.}
\begin{align}
\chi_{\ell,1}(0)&=\frac{\kappa}{\pi}\Big(K-\frac{J^2}{K}\Big)\int_0^1
\frac{(1-t^2)(1+t^2)^2}{Q_+(t)Q_-(t)}\,\dd t,\\
\chi_{\ell,2}(0)&=\frac{\kappa J^2}{K\pi}\int_0^1
\frac{(1-t^2)(1+t^2)\big[L(t^2)^2M(t^2)+4\kappa^2t^2P(t^2)\big]}{Q_0(t)A_+(t)A_-(t)}\,\dd t,\\
\chi_d(0)&=\frac{32K\kappa}{\pi}\int_0^1\frac{s(1-s^2)^2}{B_+(s)B_-(s)}\,\dd s,
\end{align}
with the auxiliary polynomials
\begin{align}
Q_\pm(t)&=(1+t^2)^2+(1\pm\alpha)^2\kappa^2t^2, & Q_0(t)&=(1+t^2)^2+\kappa^2t^2,\\
L(z)&=(\kappa+2)+(2-\kappa)z, & M(z)&=(\kappa+4)+(4-\kappa)z,\\
P(z)&=(\kappa(1+\alpha^2)+4)+(4-\kappa(1+\alpha^2))z,\\
A_\pm(t)&=L(t^2)^2+4(1\pm\alpha)^2\kappa^2t^2,\\
B_\pm(s)&=4(1-s^2)^2+\kappa^2\big[(1\pm\alpha)(1+s^2)\pm2\alpha s\big]^2.
\end{align}

\paragraph{Closed form of the first locked term.}
\begin{equation}
\chi_{\ell,1}(0)=\frac{K^2-J^2}{4\pi JK}
\Big[(K+J)\arctan\tfrac{(K+J)\kappa}{2K}-(K-J)\arctan\tfrac{(K-J)\kappa}{2K}\Big].
\end{equation}
(The prefactor $(K^2-J^2)/(JK)$ is essential; dropping it gives the wrong roots.)

\paragraph{Closed form of the sideband and drift terms.}
The other two susceptibilities are likewise finite sums of arctangents, one term per
pole of the resolvent. For $\chi_{\ell,2}$ define
\begin{align}
q_\pm&=\Big(\tfrac{\sqrt{\kappa^2+4}\pm\kappa}{2}\Big)^2,\qquad
s_{+,\pm}=\Big(\tfrac{\sqrt{4+\alpha(\alpha+2)\kappa^2}\pm\kappa(1+\alpha)}{\kappa-2}\Big)^2,
\nonumber\\
s_{-,\pm}&=\Big(\tfrac{\sqrt{4+\alpha(\alpha-2)\kappa^2}\pm\kappa(1-\alpha)}{\kappa-2}\Big)^2,
\end{align}
and collect them as $(\rho_1,\dots,\rho_6)=(q_+,q_-,s_{+,+},s_{+,-},s_{-,+},s_{-,-})$.
Then
\begin{equation}
\chi_{\ell,2}(0)=\frac{\kappa J^2}{K\pi(\kappa-2)^4}\sum_{j=1}^{6}
\frac{C_j}{\sqrt{\rho_j}}\arctan\frac{1}{\sqrt{\rho_j}},\qquad
C_j=\frac{N(-\rho_j)}{\prod_{m\ne j}(\rho_m-\rho_j)},
\end{equation}
with $N(z)=(1-z^2)\big[L(z)^2M(z)+4\kappa^2zP(z)\big]$ and $L,M,P$ the polynomials
defined above. For $\chi_d$ define
\begin{equation}
\sigma_{+,\pm}=\Big(\tfrac{\sqrt{4+\kappa^2(1+2\alpha)}\pm2}{\kappa(1+2\alpha)}\Big)^2,
\qquad
\sigma_{-,\pm}=\Big(\tfrac{\sqrt{4+\kappa^2(1-2\alpha)}\pm2}{\kappa(1-2\alpha)}\Big)^2,
\end{equation}
collected as $(\sigma_1,\dots,\sigma_4)=(\sigma_{+,+},\sigma_{+,-},\sigma_{-,+},
\sigma_{-,-})$. Then
\begin{equation}
\chi_d(0)=\frac{64K}{\pi\kappa^3(1+2\alpha)^2(1-2\alpha)^2}\sum_{j=1}^{4}
A_j\,\frac{\arctan\sqrt{\sigma_j}}{\sqrt{\sigma_j}},\qquad
A_j=\frac{\sigma_j(\sigma_j+1)}{\prod_{m\ne j}(\sigma_m-\sigma_j)}.
\end{equation}
Removable cases ($\alpha=0$, $J=K$, $\kappa=2$, $2\alpha=1$) follow by continuity. The
roots $\rho_j,\sigma_j$ may be complex (e.g.\ when $4+\alpha(\alpha-2)\kappa^2<0$); the
sums remain real and are evaluated by analytic continuation of
$(1/\sqrt{\rho})\arctan(1/\sqrt{\rho})$, summing conjugate pairs. These are the
arctangent sums quoted in the main text; $F(K,J)=0$ with them is the exact neutral
curve, validated against direct numerical quadrature in Table~\ref{tab:susc}.

\begin{table}[t]
\centering
\caption{Validation of the closed-form arctangent sums for the $m=\pm1$ susceptibilities
$\chi_{\ell,2}$ (six poles) and $\chi_d$ (four poles) against direct $40$-digit numerical
quadrature of the same integrals, evaluated on the neutral curve $K_2(J)$. The two
representations agree to machine precision, and $F=\chi_{\ell,1}+\chi_{\ell,2}+\chi_d-1$
vanishes on the curve, confirming the closed forms.}
\label{tab:susc}
\begin{tabular}{rrrrrr}
\hline\hline
$J$ & $K_2(J)$ & $\chi_{\ell2}$ & $\chi_d$ & $|\text{closed}-\text{quad}|$ & $|F|$\\
\hline
$2.5$ & $4.42276$ & $0.1312161$ & $0.3786635$ & $<10^{-40}$ & $<10^{-40}$\\
$3.0$ & $4.85454$ & $0.1739927$ & $0.2397741$ & $<10^{-39}$ & $<10^{-40}$\\
$6.0$ & $7.51036$ & $0.3135343$ & $0.0514812$ & $<10^{-40}$ & $<10^{-40}$\\
$10.0$ & $11.30504$ & $0.3878011$ & $0.0180347$ & $<10^{-39}$ & $<10^{-40}$\\
\hline\hline
\end{tabular}
\end{table}

\paragraph{Consequences.}
Near the cusp $\eps=K-4$,
\begin{equation}
J_{2}(K)=2+\tfrac54\eps-\tfrac{31}{64}\eps^2+\tfrac{12}{5\pi}\eps^{5/2}+O(\eps^3).
\label{eq:Jc2cusp}
\end{equation}
The flat branch $K_{2}=4$ ($0\le J\le2$) and the linear cusp slope $\tfrac54$ are
\emph{exact} (from the local expansion of $F$); the quadratic $-\tfrac{31}{64}$ and
half-integer $\tfrac{12}{5\pi}$ coefficients are high-precision \emph{numerical}
estimates from Puiseux extraction of $F$ ($8$ and $\sim5$ significant figures), not yet
proven. For large $J$, $K_{2}(J)=J+1+c_*/J+d/J^2+\cdots$, with $c_*=2+(2+
\arctan\tfrac12)/\pi$ \emph{exact} and $d\approx3.89$ \emph{numerical}.

\paragraph{Puiseux validation of the cusp coefficients.}
The two non-elementary coefficients in \eqref{eq:Jc2cusp} are obtained by Puiseux extraction
from the exact neutral curve itself: solving $F=0$ at high precision ($50$-digit arithmetic)
for a geometric sequence of $\eps$ and fitting the residual $J_2-2-\tfrac54\eps$ to
$c_2\eps^2+c_{5/2}\eps^{5/2}+c_3\eps^3$ over successive $\eps$-triples. The fitted coefficients
converge to the simple values $-\tfrac{31}{64}$ and $\tfrac{12}{5\pi}$
(Table~\ref{tab:puiseux}), which we adopt as the conjectured closed forms; they remain
numerical estimates ($8$ and $\sim5$ significant figures), not proven.

\begin{table}[t]
\centering
\caption{Puiseux extraction of the $J_2$ cusp coefficients from the exact neutral curve
$F=0$. A three-term fit $J_2-2-\tfrac54\eps=c_2\eps^2+c_{5/2}\eps^{5/2}+c_3\eps^3$ over
successive $\eps$-triples; the fitted $c_2,c_{5/2}$ converge monotonically to the
conjectured values $-31/64$ and $12/5\pi$ as $\eps\to0$.}
\label{tab:puiseux}
\begin{tabular}{lrr}
\hline\hline
$\eps$-triple ($\times10^{-3}$) & $c_2$ & $c_{5/2}$\\
\hline
$20,\,10,\,5$ & $-0.48356$ & $0.7368$\\
$10,\,5,\,2.5$ & $-0.48404$ & $0.7485$\\
$5,\,2.5,\,1.25$ & $-0.48424$ & $0.7555$\\
$2.5,\,1.25,\,0.625$ & $-0.48433$ & $0.7594$\\
\hline
conjectured & $-31/64=-0.48438$ & $12/5\pi=0.76394$\\
\hline\hline
\end{tabular}
\end{table}

\paragraph{Onset type.}
The crossing is a real zero eigenvalue, so the mixed branch bifurcates from the phase
wave through a pitchfork: the residual phase-translation symmetry $B\to e^{\ii\chi}B$ of the
emerging order parameter $B=W_\eta$ gives the equivariant amplitude equation
\begin{equation}
\dot B=\mu_L(K,J)\,B-g_3(K,J)\,|B|^2B+O(|B|^5),\qquad \mu_L\propto J_{2}(K)-J.
\end{equation}
Symmetry fixes the pitchfork form but not the sign of $g_3$, which we do not compute. We
instead \emph{infer} supercriticality ($g_3>0$) numerically: the mixed-branch
self-consistency and direct simulation both give a continuous, hysteresis-free onset, with
$|W_\eta|^2$ rising linearly from zero as $J\to J_2(K)^-$ ($|W_\eta|\sim\sqrt{J_2(K)-J}$,
Fig.~\ref{fig:verif-onset}). This is an inference from the order-parameter shape, not a
computed normal-form coefficient.

\section{Sync state: order parameter}\label{sec:sync-op}
The Ott--Antonsen formula $S_{\rm OA}=\sqrt{1-4K/(K^2-J^2)}$ for the sync order parameter
is exact on the decoupled line $J=0$. For $J\neq0$ it underestimates the order parameter: at
$(K,J)=(6,3)$, $S_{\rm OA}=0.333$ against $S_{\rm sim}=0.462$. It also does not give the
sync onset for $J\neq0$: its radicand vanishes at $K=2+\sqrt{4+J^2}$, not at the async
threshold $K_1=4$, which comes from the separate incoherent-state calculation. It should not be
used for finite-amplitude boundaries such as the sync$\to$mixed line $J_3$.

\paragraph{Explicit residual.} At fixed $(\nu,\mu)$ define $f_{nm}=\langle e^{-\ii(n\xi+m\eta)}\rangle$. For
$\dot\xi=\nu-Kr\sin\xi-Js\sin\eta$, $\dot\eta=\mu-Jr\sin\xi-Ks\sin\eta$ the exact moment
hierarchy is
\begin{equation*}
\dot f_{nm}=-\ii(n\nu+m\mu)f_{nm}+\tfrac{(nK+mJ)r}{2}(f_{n-1,m}-f_{n+1,m})
+\tfrac{(nJ+mK)s}{2}(f_{n,m-1}-f_{n,m+1}).
\end{equation*}
The product-Poisson ansatz is $f_{nm}=z_\xi^{\,n}z_\eta^{\,m}$ for $n,m\ge0$, with conjugates taken
coordinatewise for negative indices (e.g.\ $f_{-n,m}=\bar z_\xi^{\,n}z_\eta^{\,m}$). The $(1,0)$ and
$(0,1)$ modes give
$\dot z_\xi=-\ii\nu z_\xi+\tfrac{Kr}{2}(1-z_\xi^2)+\tfrac{Js}{2}z_\xi(\bar z_\eta-z_\eta)$ and the analogous
equation for $z_\eta$. The ansatz derivative minus the hierarchy then leaves, for $n,m\ge1$,
\begin{equation*}
R_{nm}=\tfrac{J}{2}\,z_\xi^{\,n-1}z_\eta^{\,m-1}\big[m\,r\,z_\eta(|z_\xi|^2-1)+n\,s\,z_\xi(|z_\eta|^2-1)\big],
\end{equation*}
so $R_{11}=\tfrac{J}{2}[r z_\eta(|z_\xi|^2-1)+s z_\xi(|z_\eta|^2-1)]$ and
$R_{21}=\tfrac{J}{2}z_\xi[r z_\eta(|z_\xi|^2-1)+2s z_\xi(|z_\eta|^2-1)]$ (checked symbolically, also for
modes $(1,2)$, $(2,2)$, $(3,2)$). Frequencies and self-coupling cancel from the residual.

\paragraph{Interpretation.} The full product-Poisson family is not invariant for generic
nonzero $J$, $r$ and $s$. Special invariant cases include the uncoupled case $J=0$, incoherence,
the phase-wave marginal ($s=0$), and point measures ($|z_\xi|=|z_\eta|=1$), which represent
instantaneous oscillator positions, moving or not. The original construction~\cite{Yoon2022}
includes the time-dependent unit-modulus case; its sync formula follows from a stationary
continuation outside the simultaneously locked region, which cannot reproduce the correlated
invariant measures of librating/winding orbits. The residual therefore invalidates that
stationary closure as a general order-parameter calculation, not every time-dependent
unit-modulus trajectory. For a single angle whose velocity contains only its first harmonic,
the Poisson family is invariant; the phase-wave marginal has this structure and yields
$r_{\rm pw}=\sqrt{1-4/K}$ exactly.

The full order parameter is the frequency average of the full orbit average $\overline{\cos\xi}$
over four populations,
\begin{equation}
r_{\rm sync}=\mathcal{L}+\mathcal{T}^{\xi}_{\xi\,{\rm bd},\,\eta\,{\rm w}}+\mathcal{T}^{\xi}_{\eta\,{\rm bd},\,\xi\,{\rm w}}
+\mathcal{D}^{\xi}_{\rm both\,w},
\label{eq:sync-sc-exact}
\end{equation}
where $\mathcal{L}=r_{\rm lock}$ is the both-locked contribution; the two $\mathcal{T}$ terms are the
tongue ($\xi$ bounded, $\eta$ winding) and the \emph{opposite} tongue ($\eta$ bounded, $\xi$
winding); and $\mathcal{D}$ is the both-winding population. The opposite tongue is generally
nonzero: because $\dot\xi$ depends on the coupled $\eta$ orbit, a winding $\xi$ need not have zero
mean cosine. (For example, at $K=6$, $J=3$, $S=0.46$, the orbit with $(\nu,\mu)=(4,1)$ has
$\xi$ winding and $\overline{\cos\xi}\approx0.054$.) We know of no symmetry forcing $\mathcal{D}$ to
vanish either: frequency reversal maps an orbit to $(-\xi,-\eta)$, under which $\cos\xi$ is
even. The $\xi$-bounded tongue involves the orbit average of a forced pendulum and has no known
global closed form; in the center-slaved approximation ($\overline{\cos\xi}\approx\sqrt{1-a^2}$,
ignoring the intra-cycle libration of $\xi$) it reduces to the tongue function $\mathcal{T}$. The
main text uses $S\approx\mathcal{L}+\mathcal{T}$. Its residual combines the slaving error with the
omitted opposite-tongue and both-winding populations; we do not evaluate these separately. At
$J=0$ the both-locked and $\xi$-locked/$\eta$-drifting populations together recover the
one-dimensional locked Kuramoto integral, and the approximation is exact there.

\subsection{Self-consistency and the oscillator classes}
On the sync branch $r=s=S$ (so $S\equiv r_{\rm sync}$) an oscillator with frequencies $(\nu,\mu)$ evolves on the frozen
$2$-torus
\begin{equation}
\dot\xi=\nu-KS\sin\xi-JS\sin\eta,\qquad
\dot\eta=\mu-JS\sin\xi-KS\sin\eta,
\label{eq:sync-flow}
\end{equation}
and the order parameter is fixed by
\begin{equation}
S=\langle\!\langle\cos\xi\rangle\!\rangle
=\iint h(\nu,\mu)\,\overline{\cos\xi}(\nu,\mu;S)\,d\nu\,d\mu,\qquad
h(\nu,\mu)=\frac{8}{\pi^2[(\nu+\mu)^2+4][(\nu-\mu)^2+4]},
\label{eq:sync-sc}
\end{equation}
with $\overline{\cos\xi}$ the time-average over the orbit of \eqref{eq:sync-flow}. The orbit
topology partitions the $(\nu,\mu)$ plane into the following classes:
\begin{enumerate}
\item \textbf{Both-locked}: \eqref{eq:sync-flow} has a stable fixed point
$(\xi^*,\eta^*)$, so $\overline{\cos\xi}=\cos\xi^*$.
\item \textbf{Tongue}: $\xi$ stays bounded (it librates about a slowly moving center) while
$\eta$ winds; its center-slaved contribution is $r_{\rm tongue}$.
\item \textbf{Opposite tongue}: $\eta$ bounded, $\xi$ winding. Its $\overline{\cos\xi}$ is generally
nonzero, because $\dot\xi$ depends on the coupled $\eta$ orbit.
\item \textbf{Both-winding}: both coordinates wind. No symmetry makes this class vanish:
$\cos\xi$ is even under frequency reversal.
\end{enumerate}
We evaluate the first two classes below and neglect the last two, giving the approximation
$S\approx r_{\rm lock}+r_{\rm tongue}$.

\subsection{Both-locked contribution (closed form)}
At a fixed point of \eqref{eq:sync-flow}, $\dot\xi=\dot\eta=0$ gives the linear system
\begin{equation}
KS\sin\xi^*+JS\sin\eta^*=\nu,\qquad JS\sin\xi^*+KS\sin\eta^*=\mu,
\end{equation}
with determinant $S^2(K^2-J^2)$. Writing $a\equiv\sin\xi^*$, $b\equiv\sin\eta^*$, Cramer's rule
gives
\begin{equation}
a=\frac{K\nu-J\mu}{S(K^2-J^2)},\qquad b=\frac{K\mu-J\nu}{S(K^2-J^2)},
\end{equation}
and the stable branch ($\cos\xi^*,\cos\eta^*>0$) occupies the unit square $|a|<1$, $|b|<1$.
Inverting,
\begin{equation}
\nu=S(Ka+Jb),\qquad \mu=S(Ja+Kb),\qquad
\left|\frac{\partial(\nu,\mu)}{\partial(a,b)}\right|=S^2(K^2-J^2)=AB,
\end{equation}
with $A\equiv S(K+J)$, $B\equiv S(K-J)$. The two combinations entering $h$ factorize cleanly,
\begin{equation}
\nu+\mu=A(a+b),\qquad \nu-\mu=B(a-b),
\end{equation}
so that $(\nu+\mu)^2+4=A^2(a+b)^2+4$ and $(\nu-\mu)^2+4=B^2(a-b)^2+4$. With
$\cos\xi^*=c(a)\equiv\sqrt{1-a^2}$ the both-locked contribution is the unit-square integral
\begin{equation}
r_{\rm lock}=\frac{8AB}{\pi^2}\int_{-1}^{1}\!\!\int_{-1}^{1}
\frac{c(a)}{[A^2(a+b)^2+4]\,[B^2(a-b)^2+4]}\,db\,da .
\label{eq:rlock-square}
\end{equation}
For fixed $a$ the $b$-integrand is a product of two Lorentzians centred at $b=-a$ (half-width
$\ell\equiv2/A$) and $b=+a$ (half-width $m\equiv2/B$). Partial fractions give the elementary
primitive
\begin{equation}
\Phi_a(b)=\frac{p_a}{2}\ln\frac{(b+a)^2+\ell^2}{(b-a)^2+m^2}
+\frac{q_a}{\ell}\arctan\frac{b+a}{\ell}
+\frac{r_a}{m}\arctan\frac{b-a}{m},
\label{eq:Phi-primitive}
\end{equation}
\begin{equation}
\Delta_a=[4a^2+(\ell+m)^2][4a^2+(\ell-m)^2],\quad
p_a=\frac{4a}{\Delta_a},\quad
q_a=\frac{4a^2-\ell^2+m^2}{\Delta_a},\quad
r_a=\frac{4a^2+\ell^2-m^2}{\Delta_a},
\end{equation}
so the both-locked contribution reduces to the single arcsine-weighted quadrature
\begin{equation}
r_{\rm lock}(S;K,J)=\frac{8}{\pi^2 AB}\int_{-1}^{1}c(a)\,[\Phi_a(1)-\Phi_a(-1)]\,da .
\label{eq:rlock-final}
\end{equation}

\subsubsection*{Finite closed form for $r_{\rm lock}$}
The remaining $a$-integral can be performed in closed form via the semicircle--log
kernel. The poles of $\Delta_a$ are
\begin{equation}
q\in\left\{\frac{i(\ell+m)}{2},\,-\frac{i(\ell+m)}{2},\,
\frac{i(\ell-m)}{2},\,-\frac{i(\ell-m)}{2}\right\}.
\end{equation}
For any numerator $U(a)$ define the residue weight
\begin{equation}
\omega_q[U]=\frac{U(q)}{\Delta'(q)},\qquad
\Delta'(q)=8q\bigl[8q^2+(\ell+m)^2+(\ell-m)^2\bigr].
\end{equation}
Let $R(z)=\sqrt{z-1}\sqrt{z+1}$ with $R(z)\sim z$ as $z\to\infty$ (cut on $[-1,1]$), and
define the semicircle--log kernel
\begin{equation}
H(q,z)=\int_{-1}^{1}\frac{\sqrt{1-a^2}\,\mathrm{Log}(a-z)}{a-q}\,da.
\end{equation}
With $\tau(z)=z-R(z)$, $s_q\equiv\tau(q)$, $t_z\equiv\tau(z)$, and
$B_0=-\mathrm{Log}(2t_z)-i\pi\operatorname{sgn}\Im z$, this evaluates to
\begin{align}
H(q,z)&=\frac{\pi}{2}C_0(q,z)-\frac{\pi}{4}C_2(q,z),\\
C_0(q,z)&=-\frac{2s_q}{1-s_q^2}B_0-\frac{4s_q}{1-s_q^2}\mathrm{Log}(1-s_qt_z),\\
C_2(q,z)&=\frac{2s_qt_z^2+4s_q^2t_z-4s_q^3B_0-4s_q^3\mathrm{Log}(1-s_qt_z)
+\frac{4}{s_q}\left[-\mathrm{Log}(1-s_qt_z)-s_qt_z-\frac{s_q^2t_z^2}{2}\right]}{1-s_q^2},
\end{align}
where all logarithms are continued along straight segments and $R$ uses the branch
$R(z)\sim z$ at infinity. Define the endpoint combinations
\begin{align}
L_q(c)&=H(q,-1-ic)+H(q,-1+ic)-H(q,1+ic)-H(q,1-ic),\\
A_q(c)&=\frac{1}{2i}\left[H(q,1-ic)+H(q,-1+ic)-H(q,1+ic)-H(q,-1-ic)\right].
\end{align}
Then the fully closed both-locked contribution is
\begin{equation}
\boxed{
r_{\rm lock}
=
\frac{8}{\pi^2 AB}
\sum_{q:\,\Delta_a(q)=0}
\left[
\omega_q[2a]\{L_q(\ell)+L_q(m)\}
+
\omega_q\!\left[\frac{4a^2-\ell^2+m^2}{\ell}\right]A_q(\ell)
+
\omega_q\!\left[\frac{4a^2+\ell^2-m^2}{m}\right]A_q(m)
\right],
}
\label{eq:rlock-closed}
\end{equation}
where the sum runs over the four poles of $\Delta_a$ listed above.
This finite sum (no remaining integral) agrees with the quadrature \eqref{eq:rlock-final}
to below $10^{-10}$ at all test points. At $J=0$, $A=B$ and $\ell=m$, but the integrand keeps its $a$ dependence,
$[(b+a)^2+\ell^2]^{-1}[(b-a)^2+\ell^2]^{-1}$; there one evaluates the regular quadrature
\eqref{eq:rlock-final} directly, or takes the coalescing-pole limit with that dependence
retained. Only at $a=0$ does the inner primitive reduce to
$\int(b^2+\ell^2)^{-2}db=\tfrac{b}{2\ell^2(b^2+\ell^2)}+\tfrac{1}{2\ell^3}\arctan(b/\ell)$.

\subsection{Tongue contribution}
In the tongue $\xi$ is locked while $\eta$ winds. Slaving $\xi$ to its center ($\sin\xi^*=a$),
the $\eta$-equation in \eqref{eq:sync-flow} becomes the Adler equation
\begin{equation}
\dot\eta=\Omega-KS\sin\eta,\qquad \Omega=\mu-JS\,a .
\end{equation}
On the winding branch $|\Omega|>KS$ the orbit average of $\sin\eta$ is the standard Adler
result
\begin{equation}
b\equiv\overline{\sin\eta}
=\frac{\Omega-\operatorname{sgn}(\Omega)\sqrt{\Omega^2-(KS)^2}}{KS}\in(-1,1),\quad b\neq0,
\end{equation}
inverting to $\Omega=KS(1+b^2)/(2b)$. The tongue chart is
\begin{equation}
\nu_t(a,b)=\alpha a+\beta b,\qquad
\mu_t(a,b)=\beta a+\alpha\,\frac{1+b^2}{2b},\qquad
-1<a<1,\ \ 0<|b|<1,
\label{eq:tongue-chart}
\end{equation}
with $\alpha\equiv KS$, $\beta\equiv JS$, attaching to the both-locked square at $b=\pm1$.
The Jacobian is $D_t=\beta^2+\alpha^2(1-b^2)/(2b^2)$, and the tongue contribution is
\begin{equation}
r_{\rm tongue}=\int_{-1}^{1}c(a)\,T(a)\,da,\qquad
T(a)=\int_{0<|b|<1}h(\nu_t,\mu_t)\,D_t\,db .
\label{eq:rtongue}
\end{equation}
The inner $b$-integral is elementary: with
\begin{equation}
P_+(b)=\alpha+2(\alpha+\beta)ab+(\alpha+2\beta)b^2,\qquad
P_-(b)=-\alpha+2(\alpha-\beta)ab+(2\beta-\alpha)b^2,
\end{equation}
one has $h\,D_t=N_a(b)/D_a(b)$ with
\begin{equation}
N_a(b)=\frac{128b^2[\beta^2b^2+\tfrac{\alpha^2}{2}(1-b^2)]}{\pi^2},\qquad
D_a(b)=[P_+(b)^2+16b^2][P_-(b)^2+16b^2],
\end{equation}
and the eight poles of $D_a$ (roots of $P_\pm(b)=\pm4ib$) give
\begin{equation}
T(a)=\operatorname{Re}\sum_{j=1}^{8}\frac{N_a(\rho_j)}{D_a'(\rho_j)}\,
\ln\!\frac{1-\rho_j}{-1-\rho_j},\qquad D_a(\rho_j)=0 .
\label{eq:T-sumlogs}
\end{equation}
On the measure-zero line $2J=K$, $P_-$ degenerates to linear; the limit is taken termwise.
Physically $\beta=JS$ measures the $\eta$-field's back-action on the locked $\xi$ (the
off-diagonal coupling); at $J=0$ it vanishes and the tongue chart collapses to
$\nu_t=\alpha a$, independent of $b$. The $\xi$-locked/$\eta$-drifting contribution remains
nonzero and, together with the both-locked contribution, completes the one-dimensional locked
Kuramoto integral for $\xi$.

\subsubsection*{Elliptic reduction of the outer $a$-integral}
Performing the $a$-integral first via the same semicircle Cauchy transform, the four
$a$-poles at fixed $b$ are
\begin{equation}
p_{\sigma,\epsilon}(b)
=\frac{4i\epsilon b-e_\sigma-d_\sigma b^2}{2C_\sigma b},
\qquad \sigma=\pm,\quad\epsilon=\pm1,
\end{equation}
with $C_+=\alpha+\beta$, $C_-=\alpha-\beta$, $e_+=\alpha$, $e_-=-\alpha$,
$d_+=\alpha+2\beta$, $d_-=2\beta-\alpha$. The Cauchy transform gives
$R(p_{\sigma,\epsilon})-p_{\sigma,\epsilon}=\big[\sqrt{Q_{\sigma,\epsilon}(b)}-(4i\epsilon b-e_\sigma-d_\sigma b^2)\big]/(2C_\sigma b)$,
with the square-root branch chosen so that $R(p)\sim p$ at large $|p|$ (indeed
$Q_{\sigma,\epsilon}=(2C_\sigma b)^2(p_{\sigma,\epsilon}^2-1)$), where
\begin{equation}
Q_{\sigma,\epsilon}(b)
=d_\sigma^2b^4-8i\epsilon d_\sigma b^3
+(2e_\sigma d_\sigma-16-4C_\sigma^2)b^2
-8i\epsilon e_\sigma b+e_\sigma^2.
\label{eq:quartic}
\end{equation}
This is a quartic in $b$, so the remaining $b$-integral lives on the genus-one curve
$y^2=Q_{\sigma,\epsilon}(b)$. With roots $Q_{\sigma,\epsilon}(b)=q_4\prod_{r=1}^4(b-b_r)$,
the standard Legendre reduction uses
\begin{equation}
t^2=\lambda\frac{b-b_1}{b-b_4},\qquad
\lambda=\frac{b_2-b_4}{b_2-b_1},\qquad
k^2=\frac{(b_3-b_4)(b_2-b_1)}{(b_2-b_4)(b_3-b_1)}.
\end{equation}
The remaining $b$-integral therefore has an elliptic reduction for generic parameters (in
terms of incomplete integrals $F(\phi,k)$, $E(\phi,k)$, $\Pi(n;\phi,k)$ on $y^2=Q_{\sigma,\epsilon}(b)$;
representative moduli are $k^2\approx0.371$ at $(K,J)=(6,3)$). We do not implement the explicit
Legendre primitives: in all calculations reported here the outer integral is evaluated by
one-dimensional numerical quadrature. Near the cusp $(K,J)=(4,2)$ the center-slaved
approximation $\overline{\cos\xi}=c(a)$ is not uniform, and the cusp amplitude is treated
separately in Sec.~S4.

\subsubsection*{Three-parameter generalization $T_{\alpha\beta\gamma}(a)$}
For the mixed state ($r\neq s$) the $\xi$-locked tongue has a split between the
$\xi$-locking coupling $\alpha$, the cross-coupling $\beta$, and the Adler amplitude
$\alpha/\gamma$ (where $\gamma$ is the ratio of the locked coordinate's order parameter to the
drifting one's).
The generalized chart is
\begin{equation}
\nu_t(a,b)=\alpha a+\beta b,\qquad
\mu_t(a,b)=\beta\gamma\, a+\frac{\alpha}{\gamma}\,\frac{1+b^2}{2b},\qquad
-1<a<1,\ 0<|b|<1,
\label{eq:tongue-chart-gen}
\end{equation}
with Jacobian $D_t(b)=\beta^2\gamma+\frac{\alpha^2}{2\gamma}\frac{1-b^2}{b^2}$.
Define
\begin{equation}
P_+(b)=\frac{\alpha}{\gamma}+2(\alpha+\beta\gamma)ab+\left(\frac{\alpha}{\gamma}+2\beta\right)b^2,
\qquad
P_-(b)=-\frac{\alpha}{\gamma}+2(\alpha-\beta\gamma)ab+\left(2\beta-\frac{\alpha}{\gamma}\right)b^2,
\end{equation}
so that $\nu_t+\mu_t=P_+(b)/(2b)$ and $\nu_t-\mu_t=P_-(b)/(2b)$. Then
\begin{equation}
N_{\alpha\beta\gamma}(b)=\frac{128b^2}{\pi^2}\left[\beta^2\gamma\, b^2
+\frac{\alpha^2}{2\gamma}(1-b^2)\right],\qquad
D_{a;\alpha\beta\gamma}(b)=\left[P_+(b)^2+16b^2\right]\left[P_-(b)^2+16b^2\right],
\end{equation}
and the residue-log formula gives
\begin{equation}
T_{\alpha\beta\gamma}(a)=\operatorname{Re}\sum_{j=1}^{8}
\frac{N_{\alpha\beta\gamma}(\zeta_j)}{D'_{a;\alpha\beta\gamma}(\zeta_j)}\,
\ln\frac{1-\zeta_j}{-1-\zeta_j},\qquad D_{a;\alpha\beta\gamma}(\zeta_j)=0.
\label{eq:T-gen}
\end{equation}
For $\gamma=1$ (sync, $r=s=S$) this reduces exactly to \eqref{eq:T-sumlogs}.
The three parameter assignments are:
\begin{center}
\begin{tabular}{llll}
\toprule
Contribution & $\alpha$ & $\beta$ & $\gamma$ \\
\midrule
$r_{\rm tongue}^{\rm sync}$ & $KS$ & $JS$ & $1$ \\
$r_{\rm tongueA}^{\rm mixed}$ ($\xi$-locked) & $Kr$ & $Js$ & $r/s$ \\
$s_{\rm tongueB}^{\rm mixed}$ ($\eta$-locked) & $Ks$ & $Jr$ & $s/r$ \\
\bottomrule
\end{tabular}
\end{center}
With these assignments the chart reproduces $\nu_t=Kr\,a+Js\,b$ and $\mu_t=Jr\,a+Ks(1+b^2)/(2b)$ for the $\xi$-locked tongue.

\subsection{Result, limits, and verification}
The center-slaved closed-form evaluation of the sync order parameter is the stable root
$0<S<1$ of
\begin{equation}
S=r_{\rm lock}(S;K,J)+r_{\rm tongue}(S;K,J),
\label{eq:sync-op-final}
\end{equation}
with $r_{\rm lock}$, $r_{\rm tongue}$ from \eqref{eq:rlock-final}, \eqref{eq:rtongue},
\eqref{eq:T-sumlogs}.

\emph{Decoupled limit $J=0$.} Then $\beta=0$, $\alpha=KS$, the $\eta$ dynamics decouple, and
\eqref{eq:tongue-chart} makes $\nu_t=\alpha a$ independent of $b$; the $b$-integration collapses
to the marginal of $h$ over $\mu$, which is Cauchy of half-width $2$,
$g_\nu(\nu)=2/[\pi(\nu^2+4)]$. Equations \eqref{eq:sync-op-final} then merge into the single
locked integral
\begin{equation}
S=\int_{-\alpha}^{\alpha}\sqrt{1-(\nu/\alpha)^2}\,\frac{2}{\pi(\nu^2+4)}\,d\nu
=\frac{\sqrt{\alpha^2+4}-2}{\alpha},\qquad \alpha=KS .
\end{equation}
Solving $KS^2=\sqrt{K^2S^2+4}-2$, i.e.\ $(KS^2+2)^2=K^2S^2+4$, gives $K^2S^4+4KS^2=K^2S^2$, hence
\begin{equation}
S^2=1-\frac{4}{K},
\end{equation}
the exact Kuramoto value, recovered only when the $\xi$-locked/$\eta$-drifting population is
included.

\emph{Off-diagonal.} For the sampled nonzero-$J$ sync points the center-slaved
$\mathcal L+\mathcal T$ exceeds $S_{\rm OA}$; this difference is not equal to $\mathcal T$. It is large near
the sync onset and closes deep in sync. The small-$J$ expansion
$S=\sqrt{1-4/K}+c_2(K)J^2+O(J^4)$ already exposes the discrepancy: $S$ and $S_{\rm OA}$ agree at
$J=0$ but the OA curvature $c_2^{\rm OA}=-2/(K^3\sqrt{1-4/K})$ is too steep
(e.g.\ at $K=6$, $c_2\approx-0.0107$ in the center-slaved approximation vs.\
$c_2^{\rm OA}=-0.0160$).

\emph{Verification.} Deterministic-Cauchy simulation (the documented sweep of main-text Fig.~3,
$N=2.5\times10^5$) confirms
\eqref{eq:sync-op-final} to better than $0.5\%$, and the closed forms
\eqref{eq:rlock-final}, \eqref{eq:T-sumlogs} reproduce direct quadrature of
\eqref{eq:rlock-square}, \eqref{eq:rtongue} to machine precision:
\begin{center}
\begin{tabular}{cccccc}
\toprule
$(K,J)$ & $S_{\rm OA}$ & $r_{\rm lock}$ & $r_{\rm tongue}$ & $S$ (theory) & $S$ (sim)\\
\midrule
$(6,3)$ & $0.333333$ & $0.352253$ & $0.108217$ & $0.460471$ & $0.462$\\
$(7,3)$ & $0.547723$ & $0.517574$ & $0.068465$ & $0.586038$ & $0.586$\\
$(8,3)$ & $0.646670$ & $0.618276$ & $0.046103$ & $0.664378$ & $0.664$\\
$(6,0)$ & $0.577350$ & $0.519938$ & $0.057412$ & $0.577350$ & $0.577$\\
\bottomrule
\end{tabular}
\end{center}

\subsection{Unified ordered-state order parameters}

\begin{tcolorbox}[
  enhanced,
  colback=blue!4!white,
  colframe=blue!40!black,
  title={\textbf{Swarmalator special functions}},
  fonttitle=\bfseries,
  boxrule=0.8pt,
  arc=4pt
]
\emph{Locked function} (both-locked population; for $K>|J|$):
\begin{equation*}
\mathcal L_\xi(r,s)=\frac{8rs(K^2-J^2)}{\pi^2}\int_{-1}^{1}\!\!\int_{-1}^{1}
\frac{\sqrt{1-a^2}\,da\,db}{[(K+J)^2(ra+sb)^2+4]\,[(K-J)^2(ra-sb)^2+4]},
\end{equation*}
with $\mathcal L_\eta(r,s)=\mathcal L_\xi(s,r)$. The inner integral is elementary; at $r=s$ the outer
one reduces to a finite sum of complex logarithms [Eq.~\eqref{eq:rlock-closed}].

\emph{Tongue function} (one coordinate locked, the other winding):
\begin{equation*}
\mathcal{T}(\alpha,\beta,\gamma)=\int_{-1}^1\sqrt{1-a^2}\,T_{\alpha\beta\gamma}(a)\,da,
\end{equation*}
with the inner $b$-integral a finite sum of logarithms [Eq.~\eqref{eq:T-sumlogs}]. The outer
integral is evaluated by one-dimensional quadrature; an elliptic reduction is possible but
not needed.

\emph{Drift function} (both coordinates winding):
\begin{equation*}
\mathcal{D}(r,s)=\iint_{\mathcal D_{\rm rem}}h(\nu,\mu)\,\overline{\cos\xi}\,\dd\nu\,\dd\mu,
\end{equation*}
genuinely two-dimensional (no Adler or elliptic reduction) and orbit-averaged numerically.

\begin{center}
\begin{tabular}{lll}
\toprule
State & $r$ & $s$\\
\midrule
Phase wave & $\to\sqrt{1-4/K}$ & $0$\\
Sync & $\mathcal{L}+\mathcal{T}_{\rm exact}+\mathcal{D}$ \ [$\approx\mathcal L+\mathcal T$] & $=r$\\
Mixed & $\mathcal{L}(r,s)+\mathcal{T}(r,s)+\mathcal{D}(r,s)$ & $\mathcal{L}(s,r)+\mathcal{T}(s,r)+\mathcal{D}(s,r)$\\
\bottomrule
\end{tabular}
\end{center}
\smallskip
We write $\mathcal L(r,s),\mathcal T(r,s),\mathcal D(r,s)$ for these functions with arguments
fixed by the state's $(r,s)$ (explicit coupling maps below). In the center-slaved evaluation
only $\mathcal{D}$ requires full orbit averaging. It is small: $\lesssim0.3\%$ of $r$ and up to
$\sim10\%$ of $s$ deep in the mixed wedge, and about $0.01$ at the sync point $(6,3)$.
\end{tcolorbox}

The locked and tongue constructions are not special to $r=s$; written as two reusable
\emph{swarmalator special functions} they organize the order parameters of all ordered states.
For $r\neq s$ the both-locked Cramer inversion is $\nu=Kr\,a+Js\,b$, $\mu=Jr\,a+Ks\,b$
(so $\nu\pm\mu=(K\pm J)(ra\pm sb)$, Jacobian $rs(K^2-J^2)$); the resulting integral is the
\emph{locked function} $\mathcal L$ --- the semicircle--log sum of \eqref{eq:rlock-final} with
amplitude scale and asymmetry $\rho=r/s$ factored out. The $\xi$-tongue ($\xi$ locked with
self-field $Kr$, $\eta$ winding with Adler amplitude $Ks$) has chart
$\nu_t=Kr\,a+Js\,b$, $\mu_t=Jr\,a+Ks(1+b^2)/(2b)$, Jacobian $D_t=KrKs(1-b^2)/(2b^2)+J^2rs$;
this is the \emph{tongue function} $\mathcal T$ --- the sum-of-logs \eqref{eq:T-sumlogs} now with
the three couplings $(Kr,Js,Ks)$ in place of $(KS,JS,KS)$. Both reduce to the sync forms at
$r=s$, and the fast residue forms reproduce direct quadrature to machine precision.

By the $\xi\!\leftrightarrow\!\eta$, $r\!\leftrightarrow\!s$ exchange symmetry the order
parameters obey
\begin{equation}
r=\mathcal L(r,s)+\mathcal T(r,s)+\mathcal R(r,s),\qquad
s=\mathcal L(s,r)+\mathcal T(s,r)+\mathcal R(s,r),
\label{eq:unified}
\end{equation}
where $\mathcal L$ is the both-locked function, $\mathcal T$ the center-slaved tongue function
(coordinate in question bounded, the other winding), and $\mathcal R$ the entire remainder: the
slaving error, the opposite tongue (coordinate in question winding, the other bounded), and the
both-winding population. $\mathcal R$ is defined by Eq.~\eqref{eq:unified} itself and is not an
integral over the both-winding region alone. The three ordered states are the three solution
types:
\begin{itemize}\itemsep2pt
\item \textbf{Sync} ($r=s=S$): neglecting $\mathcal R(S,S)$ gives the center-slaved
approximation $S\approx\mathcal L+\mathcal T$ [Eq.~\eqref{eq:sync-op-final}]; exchange symmetry equates
the two coordinates' remainders but does not make them vanish.
\item \textbf{Phase wave} ($s\to0$): $\eta$ is free and $\dot\xi=\nu-Kr\sin\xi$ is plain
Kuramoto, so \eqref{eq:unified} gives $r=\sqrt{1-4/K}$.
\item \textbf{Mixed} ($r\neq s$): the remainder is small and strongly asymmetric.
$\mathcal L+\mathcal T$ is a close approximation for the larger order parameter and a fair one for
the smaller:
\end{itemize}
\begin{center}
\begin{tabular}{ccccc}
\toprule
$(K,J)$ & ref.\ $(r,s)$ & $\mathcal L\!+\!\mathcal T$ $(r,s)$ & $\mathcal R(r,s)$ & $\mathcal R(s,r)$\\
\midrule
$(4.6,2.3)$ & $(0.305,0.190)$ & $(0.3049,0.1862)$ & $+0.0001$ & $+0.0038$\\
$(4.6,2.5)$ & $(0.353,0.085)$ & $(0.3526,0.0800)$ & $+0.0004$ & $+0.0050$\\
$(6.0,4.0)$ & $(0.555,0.170)$ & $(0.5563,0.1516)$ & $-0.0013$ & $+0.0184$\\
\bottomrule
\end{tabular}
\end{center}
The remainder, computed as the difference between the reference amplitude and
$\mathcal L+\mathcal T$, rides on the gap $r-s$: it is $\lesssim0.3\%$ for $r$ but reaches $\sim$10\% for
$s$, growing deeper in the wedge. It does not reduce to $\mathcal L,\mathcal T$, and the Fig.~3 mixed
branch is therefore computed by full numerical orbit averaging of every oscillator.

\paragraph{The both-winding average.}
The both-winding contribution is a time average of an observable over a flow on the
two-torus, then averaged over the disorder $h$. We evaluate it by full numerical orbit
averaging and do not provide a closed form. Single-angle averages---the one-dimensional Kuramoto
drift integral, or the Adler equation $\dot\phi=I-I_c\sin\phi$ of an overdamped Josephson
junction---are elementary; two-frequency averages are in general far harder, as for the d.c.\
voltage of an a.c.-driven junction, which is a torus rotation number~\cite{Glutsyuk2014}, and more
broadly in the theory of averaging with several fast angles~\cite{ArnoldKozlovNeishtadt,SVM,Neishtadt2014}.
We do not claim an impossibility result. Weighted Birkhoff averaging~\cite{Das2017} can converge
rapidly on sufficiently smooth quasiperiodic orbits satisfying the required arithmetic
conditions; near resonances, and for the frequency grid, numerical convergence checks are still
needed.

\section{Sync state stability}
We analyze the sync state by the same continuum linearization used for the phase wave:
perturb the density, $\rho=\rho_0+u\,e^{\lambda t}$, about the sync stationary density
$\rho_0$, insert into the continuity equation, linearize, and decompose the result in
Fourier modes. The calculation is trickier here because both rainbow coordinates
participate: on the phase wave one coordinate ($\eta$) was free while $\xi$ was
partially locked, so the drift response was an effectively one-dimensional transport
problem; on the sync branch the \emph{drifting} oscillators wind on a genuinely
two-dimensional torus, with no global first integral, so their contribution is a
two-dimensional adjoint PDE rather than a one-dimensional resolvent. We nonetheless
carry the reduction to a scalar marginal condition, evaluate the locked part in closed
form, and control the two-dimensional drift through the structure below.

\subsection{Base flow and locking diamond}
\begin{figure}[t]
\centering
\includegraphics[width=0.52\linewidth]{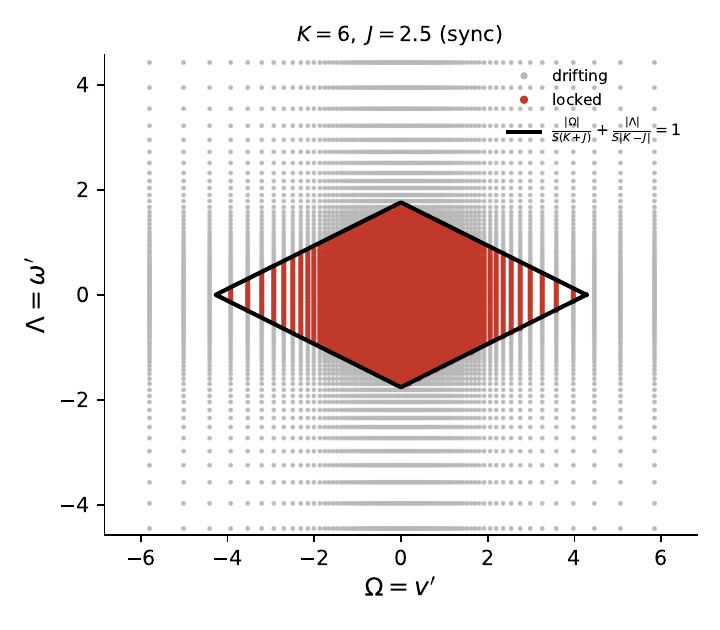}
\caption{Verification of the locking diamond \eqref{eq:sync-diamond}. Direct simulation
at a sync point ($K=6$, $J=2.5$): each oscillator is colored locked (red) or drifting
(grey) by whether $(\xi,\eta)$ stay bounded, plotted in the intrinsic-frequency plane
$(\Omega,\Lambda)=(v',\omega')$. The locked set fills the analytic diamond
(black) with no fitting.}
\label{fig:verif-diamond}
\end{figure}

On the sync branch $r=s=S$ the frozen base flow is $\dot\xi=\nu-KS\sin\xi-JS\sin\eta$,
$\dot\eta=\mu-JS\sin\xi-KS\sin\eta$. In rotated coordinates $p=\tfrac{\xi+\eta}2=x$,
$q=\tfrac{\xi-\eta}2=\theta$, frequencies $\Omega=\tfrac{\nu+\mu}2=v'$,
$\Lambda=\tfrac{\nu-\mu}2=\omega'$, and with $A=S(K+J)$, $B=S(K-J)$,
\begin{equation}
\dot p=\Omega-A\sin p\cos q,\qquad \dot q=\Lambda-B\cos p\sin q .
\end{equation}
A locked point has $\Omega=A\sin p\cos q$, $\Lambda=B\cos p\sin q$; writing
$u=\Omega/A=\sin p\cos q$, $v=\Lambda/B=\cos p\sin q$ and using $u\pm v=\sin(p\pm q)=
\sin\xi^*,\sin\eta^*$, the locking region is the diamond
\begin{equation}
|u|+|v|\le1,\qquad\text{i.e.}\quad
\frac{|\Omega|}{S|K+J|}+\frac{|\Lambda|}{S|K-J|}\le1 ,
\label{eq:sync-diamond}
\end{equation}
outside which the oscillator drifts over the $2$-torus; the stationary density splits as
$\rho_0=\rho_0^{\rm lock}+\rho_0^{\rm drift}$. Figure~\ref{fig:verif-diamond}
verifies this locking diamond directly in simulation.

\subsection{Odd-sector reduction to a scalar condition}
Linearizing the continuity equation
$\partial_t\rho+\partial_p(F_p\rho)+\partial_q(F_q\rho)=0$ about $\rho_0$ gives
$Lu=\lambda u$ with $L=U+C$: the transport
$Uu=-\partial_p(F_{p,0}u)-\partial_q(F_{q,0}u)$ [$F_{p,0}=\Omega-A\sin p\cos q$,
$F_{q,0}=\Lambda-B\cos p\sin q$] plus the finite-rank feedback $C$. The sync
state is invariant under the exchange involution
$Tu(p,q,\Omega,\Lambda)=u(p,-q,\Omega,-\Lambda)$, which commutes with $L$ and splits the
linearization into even ($Tu=u$) and odd ($Tu=-u$) sectors. Since sync$\to$mixed breaks
$r=s$ it lives in the \emph{odd} sector, where $\delta W_\xi=\ii\,Z[u]$,
$\delta W_\eta=-\ii\,Z[u]$ with the complex moment $Z[u]=b[u]+\ii\,a[u]$ and the two real
moments
\begin{equation}
a[u]=\!\int\!\sin p\sin q\,u\,\dd\Gamma,\qquad
b[u]=\!\int\!\cos p\sin q\,u\,\dd\Gamma .
\end{equation}
The feedback is rank two, $Cu=a[u]\,G_1+b[u]\,G_2$, with
\begin{align*}
G_1&=-(K+J)\,\partial_p(\rho_0\sin q\cos p)-(K-J)\,\partial_q(\rho_0\cos q\sin p),\\
G_2&=\phantom{-}(K+J)\,\partial_p(\rho_0\sin q\sin p)-(K-J)\,\partial_q(\rho_0\cos q\cos p).
\end{align*}
Because transport preserves the parity of $G_1,G_2$, the $2\times2$ characteristic
matrix is diagonal, so the odd-sector condition factorizes,
\begin{equation}
\big(1-m_a(\lambda)\big)\big(1-m_b(\lambda)\big)=0 .
\end{equation}
The $b$-channel is the odd Goldstone mode (the relative global shift of $\xi$ and
$\eta$), giving $m_b(0)=1$ identically; the nontrivial sync-breaking condition is
therefore the scalar
\begin{equation}
1-m_a(0)=0,\qquad m_a=m_a^{\rm lock}+m_a^{\rm drift}.
\label{eq:sync-scalar}
\end{equation}

\subsection{Status of the spectral argument}
Equation~\eqref{eq:sync-scalar} is the exchange-odd \emph{stationary} neutral condition. The
feedback $C=a[u]\,G_1+b[u]\,G_2$ is algebraically rank two. On a specified common operator
domain, a bounded finite-rank perturbation would preserve the appropriate essential spectrum;
we have not established the required functional-analytic setting for the combined locked and
drifting sectors, and what that spectrum is remains unsettled here. The transport generator $U$ preserves the drifting stationary density, so on the
drifting sector it is skew-adjoint in the weighted space and contributes spectrum on
$i\mathbb R$. The locked sector is different: displacements of locked point measures relax at
finite rates [already at the locked point $\xi=\eta=0$ the rates are $-S(K+J)$ and $-S(K-J)$],
and in the partially locked Kuramoto state the continuous spectrum is
$[-KR,0]\cup i\mathbb R$ rather than $i\mathbb R$ alone \cite{Mirollo2007}. A rigorous statement
would need the operator, its domain, and a compact-perturbation theorem for the combined
locked and two-dimensional drifting sectors; we do not supply these. Nor do we exclude a
complex pair crossing away from $\lambda=0$. The steady character of the tested onsets is
established numerically (Fig.~\ref{fig:verif-onset} and the cusp-slope sweeps), not by a
spectral theorem. We therefore label $F=0$ and $H=0$ as stationary neutral conditions.

\subsection{Locked contribution}
With $c_\xi=\sqrt{1-(u+v)^2}$, $c_\eta=\sqrt{1-(u-v)^2}$ over the diamond
$\mathcal D_1$ of \eqref{eq:sync-diamond},
\begin{equation}
m_a^{\rm lock}(0)=\frac{S(K^2-J^2)}{2}\iint_{\mathcal D_1}\ell(Au)\,\ell(Bv)
\Big(\tfrac1{c_\xi}+\tfrac1{c_\eta}-c_\xi-c_\eta\Big)\dd u\,\dd v .
\end{equation}
(The prefactor is linear in $S$: differentiating the locked position
$\sin\xi^*\propto1/r$ supplies a $1/r$ that cancels one power of the change-of-variables
Jacobian $\propto S^2$, leaving $S^1$; the $\tfrac12$ is from exchange symmetrization.)

\subsection{Drift contribution and the two-dimensional obstruction}
The exact drift term is a genuinely two-dimensional adjoint problem. With observable
$\psi_a=\sin p\sin q$, forcing $f_a=\big((K+J)\cos p\sin q,\,(K-J)\sin p\cos q\big)$, and
orbit-average $\bar\psi_a$ over the drift measure $\mu_{\Omega,\Lambda}$ of the $2$-torus
flow, the gauge-fixed adjoint potential $\phi_a$ solves
\begin{equation}
(\Omega-A\sin p\cos q)\,\partial_p\phi_a+(\Lambda-B\cos p\sin q)\,\partial_q\phi_a
=-(\psi_a-\bar\psi_a),\qquad \textstyle\int\phi_a\,\dd\mu_{\Omega,\Lambda}=0,
\end{equation}
and $m_a^{\rm drift}(0)=\iint_{\mathcal D_1^c}\ell(\Omega)\ell(\Lambda)
\Big(\!\int\rho_0^{\rm drift}\,f_a\!\cdot\!\nabla\phi_a\,\dd p\,\dd q\Big)\dd\Omega\,\dd\Lambda$.
Unlike Mirollo's partially locked Kuramoto problem---where
each drifter moves on a circle and the adjoint problem is a one-dimensional ODE---here
the drift flow is genuinely two-dimensional, with no global first integral, so the
adjoint equation is a first-order PDE in two variables; this is what blocks a global
closed form for the sync boundary.

\begin{figure}[t]
\centering
\includegraphics[width=0.66\linewidth]{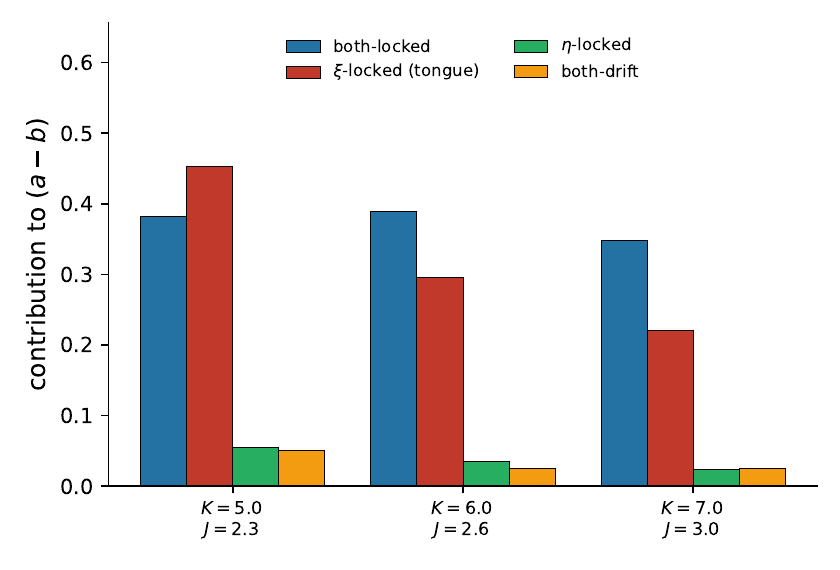}
\caption{Decomposition of the odd eigenvalue $a-b$ at the sync state by oscillator type,
at three sampled sync points. The both-locked core and the $\xi$-locked (tongue) oscillators
carry most of $a-b$; the two $\xi$-drifting channels ($\eta$-locked and both-winding) together
contribute about $8$--$11\%$, and the both-winding channel alone about $3$--$5\%$.}
\label{fig:verif-decomp}
\end{figure}

Two facts nonetheless tame the boundary near the cusp and at large $J$. \emph{(i)} At three
sampled sync points the two $\xi$-drifting response channels ($\eta$-locked and both-winding)
together contribute about $8$--$11\%$ of the total response, and the both-winding channel alone
about $3$--$5\%$ (Fig.~\ref{fig:verif-decomp}); these are response fractions, not
order-parameter contributions. The dominant drift contribution to $m_a^{\rm drift}$ at these points
comes from the $\xi$-\emph{locked}, $\eta$-\emph{drifting} tongues. This is a numerical
observation at sampled points, not a spectral exclusion argument.

\begin{figure}[t]
\centering
\includegraphics[width=0.56\linewidth]{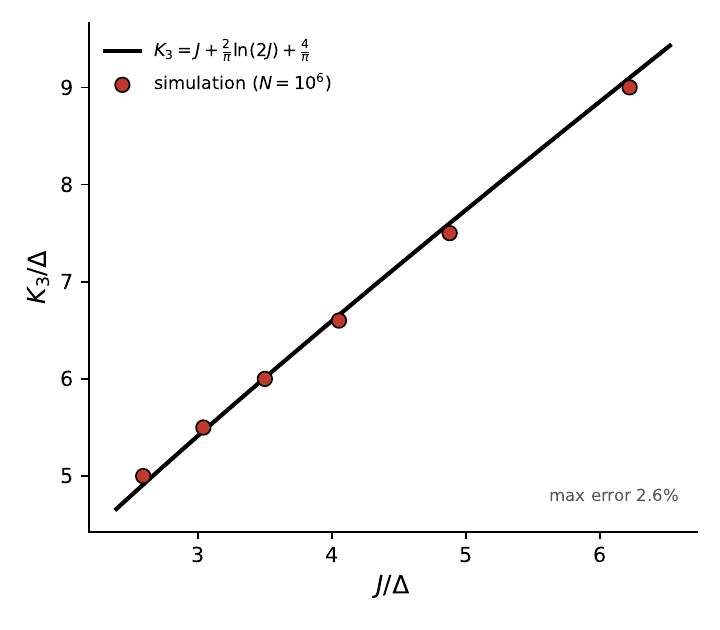}
\caption{Validation of the large-$J$ sync boundary. Markers: the sync$\to$mixed
threshold $K_{3}(J)$ from direct simulation ($N=10^6$, deterministic-Cauchy
frequencies), located as the coupling at which the sync state $r=s$ loses stability to
the mixed branch under an adiabatic sweep. Curve: the parameter-free asymptotic
$K_{3}=J+\tfrac2\pi\ln(2J)+\tfrac4\pi$, with no fitting. The maximum discrepancy is about
$2.63\%$ in $J$ relative to the inverse of the proposed curve.}
\label{fig:sync-boundary}
\end{figure}

\emph{(ii)} For those tongues $\xi$ performs a small libration driven by the winding
$\eta$, governed by a \emph{one}-dimensional adjoint resolvent on the $\eta$-orbit,
\begin{equation}
(M-Ks\sin\eta)\,w'(\eta)+\lambda_\xi\,w(\eta)=\sin\eta-\langle\sin\eta\rangle,\qquad
\lambda_\xi=Kr\cos\xi^*,\quad M=\mu-Jr\sin\xi^*,
\end{equation}
solvable in closed form by an integrating factor (it reproduces the orbit average to
$\lesssim1\%$). The marginal \emph{eigenvalue} then reduces to a one-dimensional
boundary layer at the $\eta$-locking edge $\dot\eta\to0$. This gives two asymptotic
descriptions of the boundary. At large $J$ the $\eta$-locking edge becomes a thin fast
barrier near $\cos q=0$; rescaling the barrier coordinate $y$ there, the first-harmonic
response obeys $(2+\ii y)Z_1=1$, so the in-quadrature kernel is
$\mathcal K(y)=-\Imm Z_1=y/(y^2+4)$, whose finite part
\begin{equation}
\mathrm{FP}\!\int_0^\infty\!\Big[\frac{y}{y^2+4}-\frac{\mathbf 1_{y>1}}{y}\Big]\dd y=-\log2 .
\end{equation}
A marginal balance of this kernel against the $2/\pi$ linear-response normalization of the
locked band motivates the logarithmic growth $J+\tfrac2\pi\log(2J)$, with the $-\log2$ finite
part setting the $O(1)$ offset in that balance. This motivates the proposed law
\begin{equation}
K_{3}(J)=J+\tfrac2\pi\log(2J)+\tfrac4\pi+o(1),
\end{equation}
a matched derivation with a controlled remainder is not supplied, so neither constant is
proved. We test the law numerically (Fig.~\ref{fig:sync-boundary}): over six direct $N=10^6$
deterministic-Cauchy boundary points with $J\in[2.6,6.2]$, the maximum discrepancy is about
$2.63\%$ in $J$ relative to the inverse of the proposed curve. The boundary is located by a threshold criterion on an
adiabatic sweep in steps $\Delta J=0.04$, so the quoted $2.6\%$ is dominated by finite-step
detection rather than statistical error, and the agreement does not by itself pin the two
constants independently.

Near the cusp, the slope follows from the order-parameter map conditional on two inputs
(analyzed in the next subsection): the proposed leading moment coefficients (including
$S^2\approx\eps/8$) and the inner-flow constant $\Delta H=\tfrac65$. Neither is established here,
so the resulting value $7/20$ is a conjecture. A global closed form for
$K_{3}$, valid between these limits, remains open (the
two-dimensional drift obstruction above).

\subsection{Cusp slope from the order-parameter map}\label{sec:cuspmap}
A reduction valid in a neighborhood of the cusp expresses the marginal condition
$m_a(0)=a-b=1$ through the Taylor coefficients of the frozen-field map
$\Phi_\xi(r,s)=\langle\!\langle\cos\xi\rangle\!\rangle$, bypassing the boundary layer
entirely; here $a=\partial_r\Phi_\xi$, $b=\partial_s\Phi_\xi$ at the sync point
$(S,S)$. Expanding (with $x=r^2$, $y=s^2$)
\begin{equation}
\Phi_\xi(r,s)=\tfrac{K}{4}\,r-\tfrac{K^3}{64}\,r^3-\tfrac{KJ^2}{16}\,r s^2
+r\,H(x,y)+\cdots,
\end{equation}
the diagonal coefficients are exact, following from the
decoupled cut $s=0$ ($\dot\xi=\nu-Kr\sin\xi$, $\nu$ Cauchy of width $2$, giving
$\Phi_\xi(r,0)=r-r^3+2r^5+\cdots$ at the cusp), while the cross coefficient
$c_y=-KJ^2/16$ comes from the leading-order small-field Cauchy moment hierarchy; $H$ is homogeneous of
degree two in $(x,y)$ (so $rH$ is quintic in $(r,s)$). The sync fixed point
$\Phi_\xi(S,S)=S$ fixes the amplitude,
\begin{equation}
S^2=\frac{\eps}{8}+O(\eps^2),\qquad \eps=K-4 .
\end{equation}
Subtracting the fixed-point relation from $a-b=1$ cancels the linear and
exchange-even terms order by order and leaves the compact marginal condition
\begin{equation}
(c_x-c_y)+\Delta H\,S^2=0,\qquad \Delta H:=H_x(1,1)-H_y(1,1),\label{eq:marginal-map}
\end{equation}
with $c_x=-K^3/64$, $c_y=-KJ^2/16$ the cubic coefficients and $\Delta H$ the
symmetric-point directional derivative of the quintic. Since $c_x-c_y$ vanishes at
the cusp with $\partial_K(c_x-c_y)=-\tfrac12$ and $\partial_J(c_x-c_y)=1$, and
$S^2=\eps/8$,
\begin{equation}
J_{3}(K)=2+j^*\eps,\qquad j^*=\frac12-\frac{\Delta H}{8}.\label{eq:slope-map}
\end{equation}
The quintic constant $\Delta H$ is set by the universal double-resonant inner flow at
the cusp, $\dot\xi=\hat\nu-4a\sin\xi-2b\sin\eta$,
$\dot\eta=\hat\mu-2a\sin\xi-4b\sin\eta$ (frequencies and amplitudes rescaled as
$\nu=\rho\hat\nu$, $r=\rho a$, $\dots$, so that the flow is amplitude-independent and
free of the marginal degeneracy). A direct evaluation of its exchange-odd response by
the full mean-field dynamics (a bias-cancelling antisymmetric estimator that carries the
true libration) returns a positive value of order unity in the resolvable amplitude
window, consistent with $\Delta H=6/5$ and with the opposite sign to the regular value
$\Delta H_{\rm reg}=-19/4$; the near-marginal signal at the cusp is too small for these
estimators to pin the constant to the digit. Taking $\Delta H=6/5$ gives
\begin{equation}
j^*=\frac12-\frac18\cdot\frac65=\frac{7}{20}.
\end{equation}
Conditional on the leading-order Cauchy moment coefficients (the cross coefficient and
$S^2=\eps/8$) and on $\Delta H=6/5$, the calculation gives $j^*=7/20$; neither input is
established by the present argument, so $j^*\approx7/20$ is a conjecture. A direct simulation slope cannot
adjudicate the rational: the wedge width and the symmetry-breaking amplitude both vanish as
$\eps\to0$ (the $s$-floor), so a linear fit of $(r-s)^2\to0$ is uncontrolled near the cusp. What
the simulation \emph{does} establish robustly is the sign and scale: the mixed wedge stays narrow
and persists well below the regular-hierarchy prediction $J_3=2+\tfrac{35}{32}\eps$, excluding
$\Delta H_{\rm reg}<0$ and favouring a small positive slope of order $7/20$. Analytically, $\Delta H$ is a Hadamard finite part. The regular Fourier hierarchy
gives the baseline $\Delta H_{\rm reg}=-19/4$ (slope $35/32$); the correction comes from
the resonant sectors---two primary fast-angle strips, six resonance rays
($z=\pm u,\pm2u,\pm\tfrac{u}{2}$, each reducing to a one-dimensional Adler layer with an
elementary kernel) and four locking-edge layers ($u,z=\pm4$)---with the smooth
non-resonant corner vanishing and (by a parity argument for the even frequency density)
no $\rho^4\log\rho$ term. The ray layers are individually well defined but their
distributions are ambiguous through the common frequency origin; a single intersection
counterterm, fixed by matching the double-inner core at $(\hat\nu,\hat\mu)=0$, is
required to complete the finite part---so a decomposition into rays and edges alone does
not close it. The value $\Delta H=6/5$ is the candidate consistent with this structure
and with the numerics; the closed-form intersection identity, and hence a complete proof,
is left to future work. Table~\ref{tab:coeffs} summarizes the status of all cusp and large-$J$
coefficients of the two ordered-state boundaries.

\begin{table}[t]
\centering
\caption{Status of the cusp ($\eps=K-4\to0$) and large-$J$ coefficients of the two
ordered-state boundaries. ``Exact'' coefficients follow in closed form from the
corresponding analysis of the phase-wave neutral curve $F=0$ ($J_2,K_2$ rows); ``numerical''
ones are high-precision estimates without a closed form; ``proposed'' ones come from the
large-$J$ boundary-layer balance without a controlled matched derivation and are tested
numerically; the $J_3$ cusp slope is conditional on both the proposed leading moment
coefficients (including $S^2\approx\eps/8$) and $\Delta H=6/5$, neither established here
(Sec.~S4, cusp-slope subsection).}
\label{tab:coeffs}
\begin{tabular}{lllr l}
\hline\hline
boundary & regime & coefficient & value & status\\
\hline
$J_2(K)$ & cusp & slope $\tfrac54$ & $1.2500$ & exact\\
         & cusp & quadratic $-\tfrac{31}{64}$ & $-0.4844$ & numerical\\
         & cusp & half-integer $\tfrac{12}{5\pi}$ ($\eps^{5/2}$) & $0.7639$ & numerical\\
$J_3(K)$ & cusp & slope $\tfrac{7}{20}$ \, ($\Delta H=\tfrac65$) & $0.3500$ & conjectural\\
$K_2(J)$ & large $J$ & $c_*=2+(2+\arctan\tfrac12)/\pi$ & $2.7842$ & exact\\
         & large $J$ & $d$ \, ($O(1/J^2)$) & $3.89$ & numerical\\
$K_3(J)$ & large $J$ & log coefficient $\tfrac2\pi$ & $0.6366$ & proposed; tested\\
         & large $J$ & constant $\tfrac4\pi$ (from $\mathrm{FP}=-\log2$) & $1.2732$ & proposed; tested\\
\hline\hline
\end{tabular}
\end{table}

\section{Mixed-branch self-consistency}
Freezing $(r,s)$, the locked fixed point of the frozen 2-torus flow is explicit
because the locking equations $\nu=Kr\sin\xi+Js\sin\eta$,
$\mu=Jr\sin\xi+Ks\sin\eta$ are linear in $(\sin\xi,\sin\eta)$, with determinant
$rs(K^2-J^2)$. We work in the interior mixed wedge ($r,s>0$ and $K>|J|$), where this
determinant is positive; Cramer's rule then gives
\begin{equation}
\sin\xi^*=\frac{K\nu-J\mu}{r(K^2-J^2)},\qquad
\sin\eta^*=\frac{K\mu-J\nu}{s(K^2-J^2)} .
\end{equation}
The stable fixed point is the branch $\cos\xi^*,\cos\eta^*>0$ (negative trace, positive
determinant of the linearization),
valid inside the locking band $|\sin\xi^*|,|\sin\eta^*|\le1$; oscillators outside the
both-locked band occupy tongue or both-drift classes and are orbit-averaged numerically.
The mixed branch is obtained by evaluating the combined orbit- and frequency-average,
$r=\langle\!\langle\cos\xi\rangle\!\rangle$, $s=\langle\!\langle\cos\eta\rangle\!\rangle$, on an $(r,s)$
grid and locating the crossing of the zero contours of the residual. For the curves shown in
the main text every oscillator is orbit-averaged numerically (no population decomposition),
on a $120\times120$ deterministic-Cauchy frequency grid with RK4 integration of the frozen
flow; parameters and accuracy are given in main-text Sec.~IV.C.

\section{Robustness to the frequency distribution}
\begin{figure}[t]
\centering
\includegraphics[width=0.82\linewidth]{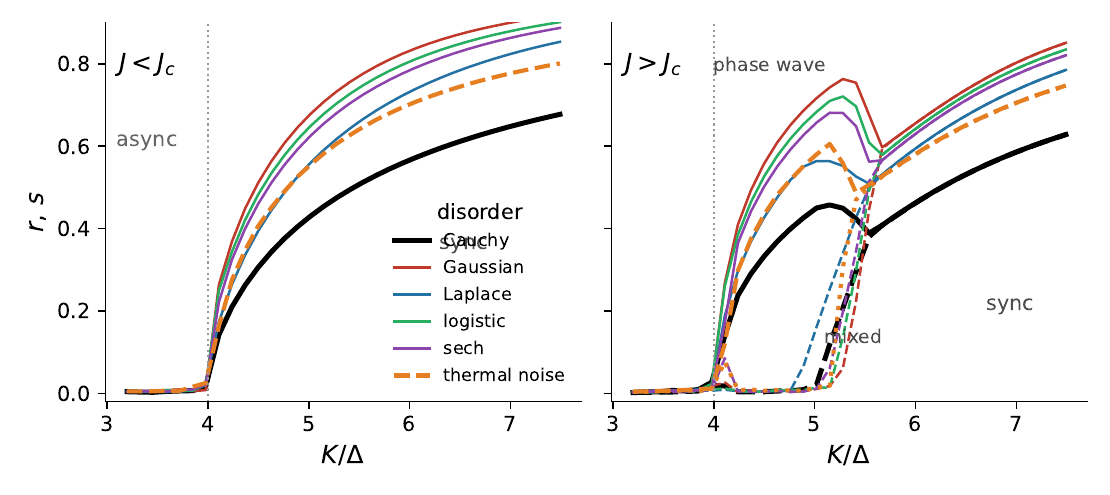}
\caption{Rainbow order parameters $r,s$ versus $K/\Delta$ for five full-support
unimodal distributions (Cauchy, Gaussian, Laplace, logistic, sech), plus thermal
noise (orange, dashed), all normalized to
$K_{1}=4$. \textbf{Left} ($J=1$): a single async$\to$sync transition.
\textbf{Right} ($J=3$): the full async$\to$phase wave$\to$mixed$\to$sync cascade
(each disorder: $\max(r,s)$ solid, $\min(r,s)$ dashed; thermal noise: $\max(r,s)$
dashed, $\min(r,s)$ dotted, orange). The same qualitative state sequences appear for each
tested distribution and for thermal noise; only the amplitudes shift (heavy-tailed Cauchy
lowest).}
\label{fig:gomega}
\end{figure}
The Cauchy density permits the explicit frequency integrations used above, but the
sampled state sequences are not tied to it. Figure~\ref{fig:gomega}
repeats the order-parameter sweeps for four other full-support, unimodal frequency
distributions---Gaussian, Laplace, logistic, and hyperbolic-secant---each rescaled so
that the incoherence threshold $K_{1}=2/[\pi g_\nu(0)]=4$ matches the Cauchy case.
The sampled $J=1$ and $J=3$ sweeps show the same qualitative state sequences (async$\to$sync
at $J=1$; async$\to$phase wave$\to$mixed$\to$sync at $J=3$) for each tested distribution; they
do not establish a common critical $J_c$ or a complete phase diagram; only the order-parameter amplitudes shift, the heavy-tailed Cauchy case
ordering most weakly in these sweeps. We exclude compact-support distributions (uniform, triangular),
whose hard frequency edges change the locking transition qualitatively. Replacing the
quenched disorder \emph{entirely} by thermal noise---identical oscillators
($v'=\omega'=0$) with Gaussian white noise added to $\xi,\eta$, calibrated to the same
$K_{1}=4$ (orange, Fig.~\ref{fig:gomega})---reproduces the cascade as well. The
cascade is thus robust across the tested full-support distributions and thermal noise,
suggesting it is controlled by the coupling rather than the detailed disorder; we do not
claim universality beyond this tested family (compact-support disorder is excluded).